\documentclass[aps,prl,twocolumn,groupedaddress,floatfix,showkeys]{revtex4-1}
\usepackage{graphicx,color}
\usepackage{amsmath,amssymb,latexsym}
\usepackage{mathtools}
\usepackage[english]{babel}
\usepackage{dcolumn}
\usepackage{bm}
\usepackage{float}
\usepackage[normalem]{ulem}
\usepackage{color}
\usepackage{chngcntr}

\newcommand\sifigures{%
    \setcounter{figure}{0}
    \makeatletter 
       \renewcommand{\thefigure}{SI.\arabic{figure}}
    \makeatother}
\begin{document}
\author{L.~Tubiana}
\email{luca.tubiana@unitn.it}
\affiliation{Physics Department, University of Trento, via Sommarive, 14 I-38123 Trento, Italy}
\affiliation{INFN-TIFPA, Trento Institute for Fundamental Physics and Applications, I-38123 Trento, Italy}
\affiliation{Faculty  of  Physics,  University  of  Vienna,  Boltzmanngasse  5,  1090  Vienna,  Austria}

\author{F.~Ferrari}
\affiliation{CASA* and Institute of Physics, University of Szczecin, Wielkopolska 15, 70-451 Szczecin, Poland}
\author{E.~Orlandini}
\affiliation{Department of Physics and Astronomy, University of Padova, 
Via Marzolo 8, I-35131 Padova, Italy}
\affiliation{INFN, Sezione di Padova, Via Marzolo 8, I-35131 Padova, Italy}

\date{\today}

\title{Ring-o-rings: a new category of supramolecular structures with topologically tunable properties}
\begin{abstract}

Macrochains of topologically interlocked rings  with unique physical properties have recently gained considerable interest in supramolecular chemistry, biology, and soft matter. Most of the work has been, so far,  focused on linear chains and on their variety of conformational properties compared to standard polymers. Here we go beyond the linear case and show that, by circularizing such macrochains, one can exploit the topology of the local interlockings to  store torsional stress in the system, altering significantly its metric and local properties. Moreover, by properly defining the twist (Tw) and writhe (Wr) of these macrorings we show the validity of a relation equivalent to the C\v{a}lug\v{a}reanu-White-Fuller theorem $Tw + Wr$=const, originally proved for ribbon like structures such as ds-DNA. Our results suggest  that circular structures of topologically linked rings with storable and tunable torsion can form a new category of highly designable multiscale structures with potential applications in supramolecular chemistry and  material science.
\end{abstract}
\date{\today}

\maketitle
Topologically constrained molecules and polymers have recently attracted considerable attention in physics, chemistry and biology~\cite{orlandini2021topological}.  Examples are nano-engineered Mechanically Interlocked Molecules (MIMs) such as rotaxanes, catenanes, and molecular knots~\cite{lewis2017properties,liang2020dynamism}, melts of rings~\cite{kapnistos2008unexpected}, and olympic gels~\cite{de1979scaling, fischer2015formation, krajina2018active} such as the natural occurring kinetoplast DNA~\cite{chen1995topology,klotz2020equilibrium,soh2020deformation,polson2021flatness}. Steady advancements in chemical synthesis techniques~\cite{chen2011directed,wu2017poly,datta2020self}, modelling, and simulations~\cite{marenda2018discovering,ubertini2021computer} have recently started to offer a framework to design systems of interlocked rings with controllable properties~\cite{livoreil1994electrochemically,hu2015switchable,wu2017poly,datta2020self} and versatile applications~\cite{aucagne2007catalytic,ayme2015strong,liu2016weaving}. Examples range from catalyzers and nano-machines~\cite{nguyen2005reversible,leigh2003unidirectional,hernandez2004reversible,SAUVAGE2010315,erbas2015artificial} to candidates for novel smart materials and artificial muscles~\cite{sauvage2017chemical}.

Novel experimental techniques are now opening the possibility to synthesize high-weight polycatenanes, called Mechanically Interlocked Polymers (MIPs), long chains composed by $n$ elementary rings held together only by topological interlocking~\cite{wu2017poly,mena2019mechanically, datta2020self}. In particular, in~\cite{wu2017poly} metal supramolecular polymers were used to obtain polycatenanes composed of up to $n=130$ rings, while in~\cite{datta2020self}  a novel self-assembly technique was proposed in which supramolecular rings were grown directly to form a catenane of up to 22 units~\cite{datta2020self}. These studies have prompted the question of how the configurational properties of MIPs differ from their standard polymeric counterparts whose elementary units are held together by covalent bonds~\cite{rauscher2018topological,rauscher2020dynamics,dehaghani2020effects,lei2021dimensional}. 

Interestingly, in~\cite{wu2017poly} it was also demonstrated the possibility to synthesize cyclic polycatenanes. This brings up the question on how and to which extent the imposed circular constraint can affect the  physical properties of these supramolecular structures. 
For instance,  circularized polycatenanes may assume supercoiled configurations as in dsDNA, significantly affecting their elastic and dynamical properties, and, arguably, their responsiveness to external stimuli. 

Here we show that circular polycatenanes can store supramolecular torsion upon circularization and, by using extensive molecular dynamics simulations, we characterize their equilibrium  properties as a function of the number of elementary rings $n$ and the amount of supramolecular torsion trapped into the system. Our results show that this controls the average extension of the macrochains as well as their local properties such as  the relative orientation of the rings. Finally, by extending to the present case the notion  of twist and writhe  used in  ribbon-like structures such as the ds-DNAs, we show that a relation equivalent to the C\v{a}lug\v{a}reanu-White-Fuller theorem holds also for circular polycatenanes.

\begin{figure}[ht]
    \centering
    \includegraphics[width=0.9\columnwidth]{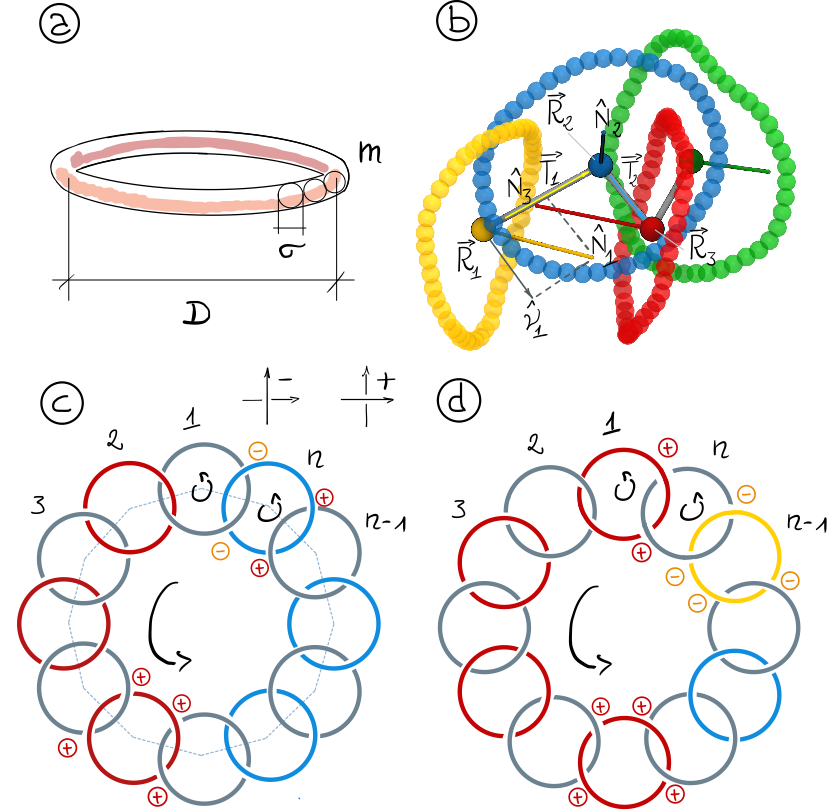}
    \caption{a) The annular polycatenanes  we consider are composed of $n$ identical, almost rigid rings of diameter $D$, each formed by $m$ beads of diameter $\sigma$. b). The stiffness of the rings allows us to map them on their centers $\mathbf{R}_i$ and normal vectors $\mathbf{\hat{N}}_i$. Notice how the vectors $\mathbf{\hat{N}}_i$ need not to be perpendicular to the backbone vectors $\mathbf{T}_i$. c) A polycatenane with torsion index $n_{tr}$ can be obtained by using $n_{tr}$ "torsion inducing" rings (red) and $n/2-n_{tr}$ "freely-rotating" rings (blue).  d) Equivalent polycatenanes with the same value of $n_{tr}$, like the one shown here, can be obtained by flipping and reorienting one or more rings. }
    \label{fig:system}
\end{figure}

Our reference system consists of $n$ semiflexible oriented rings (the elementary units of the polycatenane) each composed of $m=48$ beads with nominal diameter $\sigma$. This level of polymerization has been chosen to obtain elementary rings with a thickness to diameter ratio, $p=\frac{\sigma}{D}$, similar to the one typically achieved in  polycatenanes obtained from metal-supramolecular polymers~\cite{wu2017poly}. Other values of $p$, compatible with DNA minicircles~\cite{liang2020dynamism} and supramolecular toroids~\cite{datta2020self} have been investigated too. The connectivity of each elementary ring is provided by a FENE potential while the ring's nominal persistence length is set to $ l_p = 2m\sigma$, preventing  substantial variations in its local curvature due to thermal fluctuations (see Fig.~\ref{fig:system}a). This level of rigidity is compatible with those of the rings used both in~\cite{wu2017poly} and ~\cite{datta2020self}.
The excluded-volume interaction among the $nm$ beads  is treated via a Weeks-Chandler-Andersen (WCA) potential, see SM. 

We consider closed polycatenanes formed by $n=2k$ rings, connected  in such a way that they can always be arranged into a planar circle. Adopting this configuration as a reference, we orient the rings so that all their normals point either above or below the plane. Specifically, for each ring $i$ we define its normal as $\mathbf{N}_i = \frac{4}{m}\sum_{k=1}^{m/4} (\mathbf{r}_{k+m/4} - \mathbf{R}_i)\times(\mathbf{r}_{k} - \mathbf{R}_i)$, where $\mathbf{r}_k$ is the position of the $k-$th bead on ring $i$, oriented as above, $\mathbf{R}_i$ is its center of mass, and $\mathbf{\hat{N}}_i=\mathbf{N}_i/|\mathbf{N}_i|$. We then assign a linking number $Lk_{i,i+1}$, to all pair of rings $i, i+1$ (with $n+1=1$) according to the standard convention for the sign of crossings, see Fig.~\ref{fig:system}. In this way one can  distinguish three classes of rings forming the circular polycatenanes: $0$-rings, which form a $+1$ and a $-1$ Hopf link with their neighbours (in blue in Fig.~\ref{fig:system}), and $+2$($-2$) rings, which contribute two $+1$($-1$) Hopf links (in red and yellow respectively in Fig.~\ref{fig:system}c), d)). Clearly, the linking numbers of all ring pairs can be univocally identified by fixing the type of either even- or odd-numbered rings. Furthermore, in the planar conformation, $0$-rings can rotate around the axis identified by the centers of their neighbours along the chain by an angle $\phi(p)$ without affecting their neighbours, while the same is not possible for $+2$ and $-2$ rings. Thus, the latter can be thought to induce a torsion in the polycatenane.

Using the setting above, we identify the supramolecular torsion captured by circularization with $n_{tr} = n_{+} - n_{-}$ where $n_{+}$ and $n_{-}$ are the number of  $+2$ (red)  $-2$ rings (yellow) respectively. Clearly, $n_{tr} = \frac{1}{2} \sum_{i=1}^n Lk_{i,i+1}$, where $Lk_{i,i+1} \in \{-1,+1\}$ is the linking number of the pair of rings $(i,i+1)$, with $n+1 = 1$. Given the constraint that all polycatenanes must admit a planar conformation as in Fig.~\ref{fig:system} c), $n_{tr}$ can only take values between $-n/2$ and $n/2$.

We note that for circular polycatenanes $n_{tr}$ is conserved, while the number and position of $0$, $+2$ and $-2$ rings can vary. This can be easily seen by rotating the gray ring marked $1$ in Fig.~\ref{fig:system}c) and then reorienting the rings so that their normals point up. The resulting catenane, shown in Fig.~\ref{fig:system}d), still has the same value of $n_{tr}$, no matter whether we count it on the odd or even rings.  As shown in the SM, all different labelings of the rings giving a fixed value of $n_{tr}$ can be mapped to a reference polycatenane having $n_{tr}$ $+2$-rings and $n/2-n_{tr}$ $0$-rings.

The system is evolved  with an underdamped Langevin dynamics  integrated numerically with the LAMMPS package~\cite{plimpton1995lammps} with default values for the mass, temperature, and energy coefficients, damping time $\tau_{d}=10\tau_{LJ}$ where $\tau_{LJ}$ is the characteristic simulation time, and  integration time step $\Delta t = 0.0124 \tau_{LJ}$, see SM. Starting from an initial condition with a fixed value of $n_{tr}$, the system is relaxed to equilibrium where an extensive sampling of the configurational space is performed. Here we consider polycatenanes composed of $n=\left\{20,40,60,80,100,200,300\right\}$ rings, with $n_{tr}/n = \left\{0,0.1,0.2,0.3,0.4,0.5\right\}$.

\begin{figure}[ht]
\includegraphics[width=0.8\columnwidth]{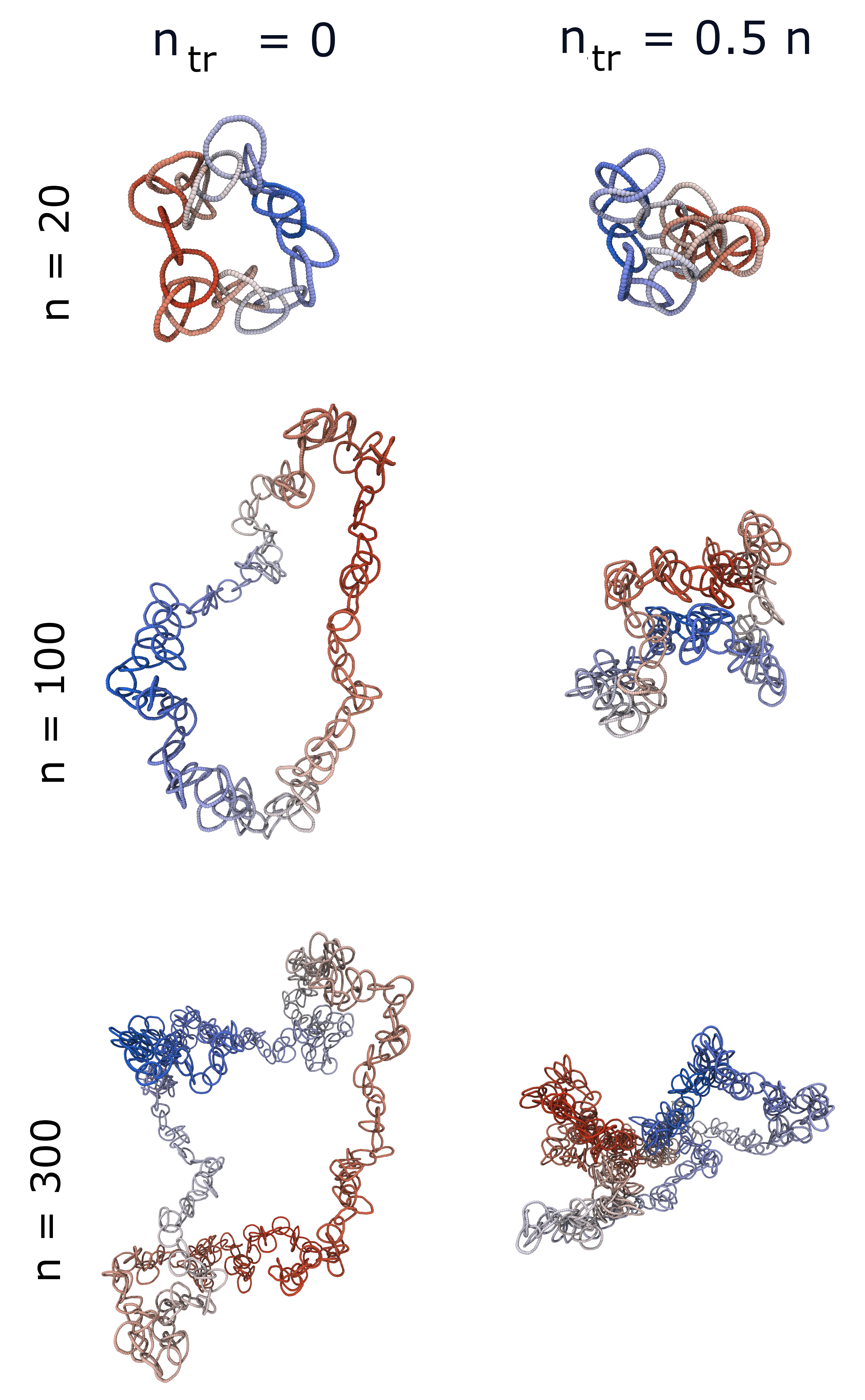}
\caption{Typical configurations for $n=20, 100, 300$ for two different values of $n_{tr}$. Left column: $n_{tr}=0$, the polycatenanes are torsionally relaxed. Right column, $n_{tr} = 0.5 n$, corresponding to the maximum torsion storable in planar circles. These last polycatenanes are visibly more compact and crumpled. The scale of the snapshot is preserved at fixed values of $n$.  The color map highlights the sequence of the elementary rings along the backbone.
}
 \label{fig:snapshot}
\end{figure}

\label{sec:results}
In figure~\ref{fig:snapshot} we report some typical equilibrium configurations of circular polycatenanes for several values of $n$ and different amount of $n_{tr}$: one can readily see that, for fixed $n$, configurations with large $n_{tr}$  are more crumpled than those that are  torsionally relaxed ($n_{tr}=0$). 
Moreover, as $n$ increases, torsionally stressed  configurations start to form 
curled substructures reminiscent of the plectonemes in supercoiled DNA~\cite{krajina2016large}. Although 
$n$ is not very large, branched-like structures at very large scale can be observed too.


\begin{figure}[ht]
    \centering
    \includegraphics[width=0.8\columnwidth]{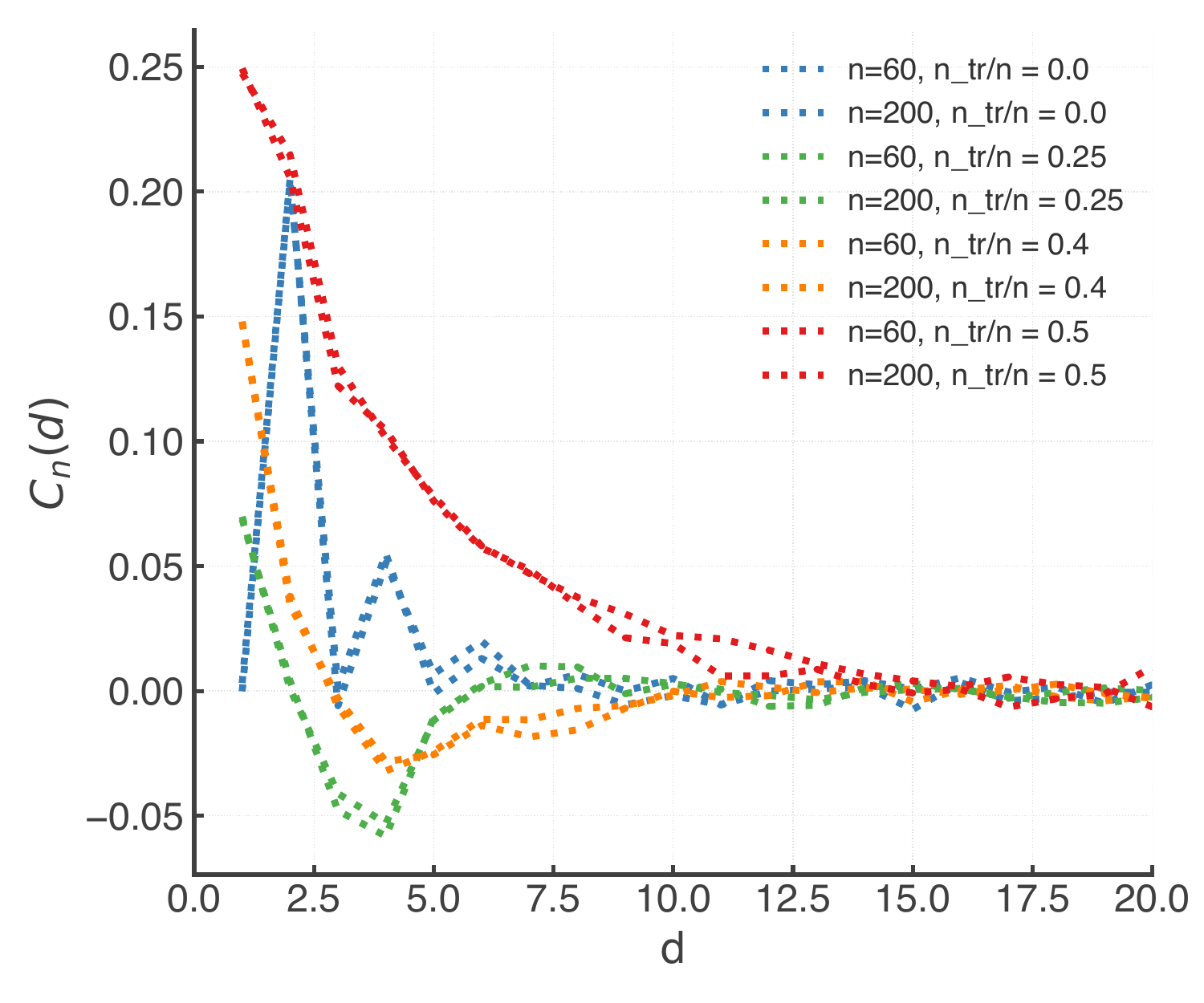}
    \caption{Normal-normal correlation function, $C_n(d)$ as a function of the index-distance $d$ along the backbone of the polycatenane, for four different values of the fraction, $n_{tr}/n=(0,0.25,0.4,0.5)$ with $n=60$ and $n=200$. Each ring is made by  $m=48$ beads.\label{fig:local}}
\end{figure}
Note that the  configurational space available in macroring models is more complex and richer than the one in twistable (or ribbon-like) chain models typically used to describe dsDNA. In fact, our "monomers" are  rings linearly bonded through topological constraints (Hopf links). This allows a significant degree of freedom for each ring, a local entropy, while keeping the whole system  globally constrained. The consequences can be appreciated by looking at the normal-normal correlation function $C_n(d)=\langle\frac 1n\sum_{i=1}^n \mathbf{\hat{N}}_i\cdot\mathbf{\hat{N}}_{i+d}\rangle$ for different values of the stored torsion, see Fig.~\ref{fig:local} a). As expected, for the maximum value of storable torsion, $n_{tr}=n/2$, $C(d)$  decreases slowly, confirming that some twist is stored  along the catenane. Interestingly, for $n_{tr}=0$  the correlation $C(d)$ goes to zero at $d=1$, and then regularly oscillates between relatively large, but decreasing values for $d=2,4,6$ and zero for $d=3,5,7$. This behaviour can be understood by observing that in this case, each ring  can freely rotate around the axis joining its two neighbours. Hence, the angle between the normals of two contiguous rings may range from $0$ to $180$ degrees giving an average of  $90$ degrees.
This means that on average  consecutive rings will lie on reciprocally orthogonal planes and that next nearest neighbours are likely to have the same orientation, i.e. $C(d)=0$ for odd $d'$s and $C(d)>0$ for even $d'$s  (the positive sign is due to our  choice of the normals in the reference configuration). Clearly, as the main direction of the backbone is changing, this peculiar even-odd behaviour tones down and disappears as $d$ becomes sufficiently large.
For $n_{tr} = \frac{n}{4}$, we see yet another set of relative orientations, with $C(d)$ reaching negative values. This remains true for $n_{tr} = 0.4 n$, although the minimum is less marked. This effect appears to be due to the emergence of local conformations  such as the one depicted in Fig.~\ref{fig:system}b), in which rings tend to stack together along the backbone. Such local organizations are no longer relevant when $n_{tr} = \max(n_{tr}) = n/2$. In this case the rotation of each ring around the catenane's backbone affects the position of the others.
\begin{figure*}[ht]
    \includegraphics[width=0.8\textwidth]{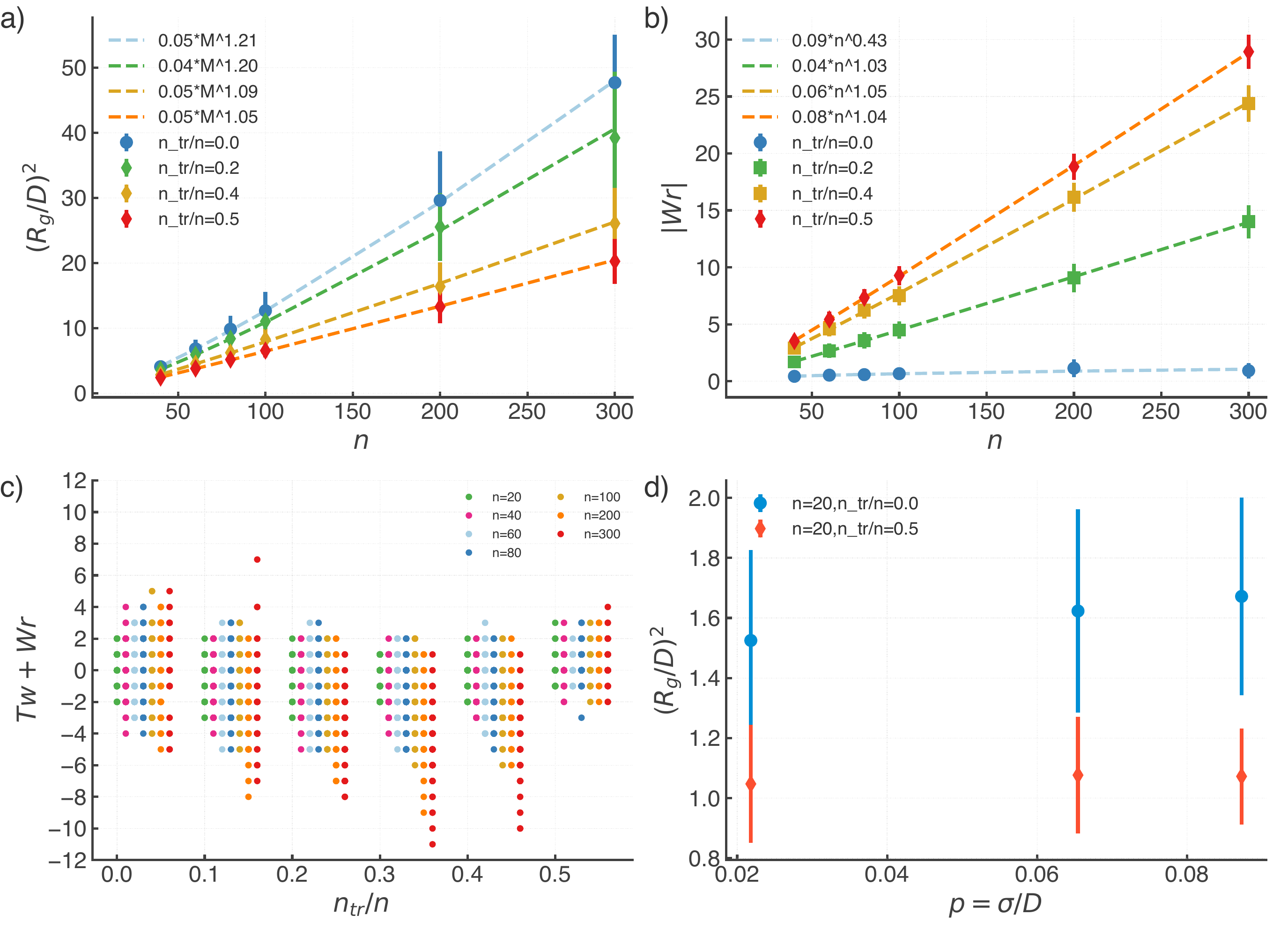}
    \caption{\label{fig:global} a) Squared radius of gyration as a function of the number of rings, $n$, for different values of $n_{tr}/n$. $R_g$ is in units of squared diameter of an ideal ring with $m$ beads, $D=m/\pi$.  b) Scaling of the absolute value of the writhe, $|Wr|$,  with $n$, for different values of $n_{tr}/n$. c) $Tw + Wr$ for different values of $n_{tr}/{n}$ and $n$. Results for different contour length $n$ have been shifted on the abscissa for clarity. Results for $n=20$ include different aspect ratios $p$. d) Dependency of $(R_g/D)^2$ on the aspect ratio of the rings, $p=\sigma/D$, for polycatenanes with $n=20$, having $n_{tr}/n = 0 $ (upper point) and $n_{tr}/n=0.5$ (lower points). The error bars in panles a), b), d) correspond to the standard deviation of the reported values.}
\end{figure*}

To  characterize quantitatively the effects that different amounts of locked torsional stress  have on the configurational properties of the system, we coarse grain it by identifying  each oriented ring $i$ with its center of mass $\mathbf{R}_i$ and its normal $\mathbf{\hat{N}}_i$. The resulting polygonal curve is  described by the sequence of pairs of vectors $(\mathbf{R}_i,\mathbf{\hat{N}}_i)$ and bonds $\mathbf{T}_i = \mathbf{R}_{i+1}-\mathbf{R}_i$, see SM for more details. 
Using this coarse-grained coordinates, we computed  the squared radius of gyration of the backbone  for different values of $n$ and $n_{tr}$: $R_g^2(n,n_{tr})$. All distances were measured in units of the ideal ring diameter, $D=\frac{m}{\pi}$. As shown in Fig.~\ref{fig:global} a),  the stored torsion affects significantly  the value of $R_g^2(n,n_{tr})$: going from $n_{tr} = 0$ to $n_{tr}=\frac{n}{2}$ at fixed $n$ the polycatenanes become roughly twice as compact.
Furthermore, despite the fact that the simulated values of $n$ are not sufficiently large to draw any conclusion on the exponent of the expected scaling behavior $R_g^2\sim A n^{2\nu}$, the different slopes of the curves $R_g^2(n,0)$ and $R_g^2(n,n/2)$ suggest that introducing a torsion above a certain level  should at least change the amplitude $A$. Note that in very long dsDNA rings, the scaling behavior of the average extension crosses over to the one expected for branched polymers  when supercoiling becomes relevant~\cite{krajina2016large}. We expect that a similar crossover occurs also  for catenanes made by a very  large number of elementary rings, a condition that, however, is currently not experimentally accessible. 

The fact that circular polycatenanes take on conformations which resemble a DNA with plectonemes (see also the snapshots reported in Fig.~\ref{fig:snapshot} for $n=300$) can be further tested by looking at the $n$ dependence of the absolute value of the writhe $|Wr|$, a quantity that captures the amount of coiling of a closed curve on itself~\cite{moffatt1995helicity,white1969self,bates2005dna}. In our case, $Wr$ is measured on the backbone of the polycatenane, defined as the polygonal curve interpolating the centers of mass of the rings. 

The behaviour of $|Wr|$ as a function of $n$ is reported in  Fig.~\ref{fig:global} b) for different values of the stored torsion. The difference in the scaling behaviour of $|Wr|$ for relaxed and torsionally stressed polycatenanes is evident.  For the former case, $|Wr| \sim n^{1/2}$, as expected by rigorous and numerical results on unconstrained random polygons~\cite{van1993writhe,portillo2011mean}. On the other hand, as soon as a finite density of torsion is trapped along the circular polycatenanes, i.~e. $n_{tr} >0$,  $|Wr|$ grows linearly with $n$ with an amplitude that depends on $n_{tr}$. This transition from  sub-linear to linear regime is a genuine effect of circular polycatenanes  that cannot be observed in standard models of linear unstructured polymer chains unless they are  either strongly confined~\cite{panagiotou2010linking,marko2011scaling,micheletti2006knotting}  or collapsed into globular shapes by effective attractive interactions~\cite{baiesi2009interplay}  

The physics of double stranded polymers as dsDNAs offers again a useful framework to characterize the conformational properties  of circular polycatenanes. In the case of a dsDNA  we know that circularization fixes  the linking number between the two strands of the helix, $Lk$, via the formula $Lk = Tw + Wr$, where $Tw$ is the total twist of the dsDNA helix around the ring backbone and the writhe $Wr$ is the average amount of coiling of the ring on itself, as stated above~\cite{fuller1978decomposition,white1969self,bates2005dna}. This result is the famous C\v{a}lug\v{a}reanu-White-Fuller theorem. For dsDNA, the definitions of $Lk$, $Tw$, and $Wr$, arise naturally from an expansion of the Gauss linking number integral~\cite{moffatt1995helicity,calugareanu1959integrale,kamien2002geometry}. This is true in general for ribbon-like surfaces, if one maps the boundaries of the ribbon to two curves linked together. 

In our case, the assumptions used to expand the Gauss linking number in the dsDNA case do not hold,  as  circular polycatenanes are not smooth; on the contrary, the local entropy of the rings can cause abrupt changes in the local properties of the macrochains.
Nonetheless, since we are considering almost rigid rings,  we can still define $Tw$ as the amount of twist captured by the normals of the rings, being careful to consider only the component orthogonal to the backbone (see SM and Fig.~\ref{fig:system}b), and  $Wr$ as the total writhe of the catenane backbone. 
By doing this, we are implicitly mapping the macroring onto a ribbon. One of the curves defining that ribbon is provided by the coarse-grained backbone discussed earlier. The other curve connects the tips of the vectors $\hat{\nu}_k$ that are obtained by projecting the normals  $\mathbf{\hat{N}}_k$ onto the direction that is perpendicular to the vectors $\mathbf t_k$ forming the backbone. In this way, we can measure $Tw$ and $Wr$ for any circular polycatenane, and check whether and to what extent the relation $Tw+Wr=constant$ holds.

As shown in Fig.~\ref{fig:global} c), we find that $Tw + Wr \in \mathbb{Z}$ for all configurations. As the twist angle between two consecutive normals is defined in $[-\pi,\pi]$, and our coarse-graining is equivalent to a piece-wise ribbon in which two successive normals can be twisted by an angle larger than $|\pi|$, this is equivalent to say that $Tw\,\textrm{mod}(1) + Wr = 0$. 
We verified this equivalence by taking its fractional part, and found it to hold up to a factor of order $10^{-10}$, with a standard deviation of $10^{-12}$, compatible with numerical round-off errors.
The fact that $Tw$ is defined $\textrm{mod(1)}$ is a known aspect of the original C\v{a}lug\v{a}reanu theorem which can be dispensed away for smooth ribbon, but might be present in general~\cite{moffatt1995helicity}.

Finally, we investigate how the configurational properties of the circular polycatenanes depend on the thickness-to-diameter ratio, $p=\sigma/D$,  of the elementary rings. This has been done in the case of rings with $n = 20$ and  $p = 0.022$,$0.65$ and $0.87$.

The results, reported in Figs.~\ref{fig:global}c) and d) with $n = 20$ and twist densities $n_{tr} = 0, 10$, clearly
show that $Tw\,\textrm{mod(1)} + Wr =0$ regardless of the thickness of the rings, as expected. Furthermore, while the average extension of the system might depend on  $p$ in a non trivial way, as larger values of $p$ correspond to a larger amount of twist which  relaxes into writhe, we notice that, for the range of values $p$ commonly used in self-assembled polycatenanes,  the observed effective compression of the system is notable and persists over about one order of magnitude. 

In conclusion, in this study we demonstrated  how the circularization of properly designed polycatenanes  allows the storage of a given amount of torsional stress that radically affects the equilibrium properties of these supramolecular structures, both locally and globally. Specifically, we identify a topological parameter, $n_{tr}$ which quantifies the amount of torsion initially stored into the polycatenane, and show that it controls  both the relative orientation of nearest-neighbour rings and the scaling of $R_g$ and $|Wr|$. 

Remarkably, our results show that the C\v{a}lug\v{a}reanu-White-Fuller relation holds for circular polycatenanes at least up to a factor $\textrm{mod(1)}$ in the definition of the twist: $Tw\, \textrm{mod(1)} + Wr = 0$ for all polycatenanes. Taken together with the result that $|Wr|$ grows as a function of $n_{tr}$, as shown in Fig.~\ref{fig:global}b), this suggest that it should be possible to map the polycatenane to a ribbon for which $Tw + Wr = Lk$, where $Lk$ is an appropriate linking number. Consider for example an open, linear polycatenane with alternating links ($n_{tr}=0$) and all normals pointing up. Clearly, one can twist its backbone by $\pi$ by simply flipping the last ring or, more in general, all rings after a chosen one. If one then circularizes the polycatenane and reassigns the normals and linking numbers, the flipped ring will correspond to either a $+2$ or $-2$ ring, $n_{tr}=\pm 1$ (see SM for more details). Therefore, it is tempting to conclude that $\pm 2$ rings correspond to a $\mp \pi$ twist of an equivalent ribbon and one crossing of its two boundaries. Then, if one considers the $Tw$ given by the original normals, which would now alternate, one would get the relation $Tw + Wr = \frac{1}{2}n_{tr}$. In general though,
due to the local entropy of the rings, mapping any given polycatenane configuration to an equivalent ribbon is challenging and requires further study.

Since our results remain valid for polycatenanes in good solvent conditions and over a wide range of ring thickness-to-diameter ratios, we believe they should be observable experimentally in systems ranging from DNA polycatenanes ~\cite{schmidt2011construction}, synthetic polymeric polycatenanes~\cite{wu2017poly,datta2020self}, to at much larger scales, macroscopic systems in which thermal fluctuations are replaced by randomized mechanical stimuli. 

Finally, we believe that the model and findings  presented here could be of interest for further developments in  supramolecular chemistry  and in the physics of soft materials, particularly soft-robotics~\cite{castano2014smart, xiong2021functional}. Moreover, as the C\v{a}lug\v{a}reanu-White-Fuller relation links a local geometrical property, the twist, to a global one, the Writhe, our results suggest that circular chains of  topologically interlocked rings  could be an inspiring  system for mathematicians and theoretical physicists ~\cite{white1969self,duplantier1981linking,moffatt1995helicity,dennis2005geometry} and be exploited to build highly responsive materials with tunable properties. 

\begin{acknowledgments}
The authors acknowledge the contribution of the COST Action Eutopia, CA17139. LT acknowledges support from the MIUR grant Rita Levi Montalcini. The research of FF and LT has been supported in part by the Polish National Science Centre under grant no. 2020/37/B/ST3/01471.
The authors are grateful to R. Ricca and T. Tarenzi for their insight in several useful discussions. 
\end{acknowledgments}

\bibliography{biblio}

\begin{thebibliography}{52}%
\makeatletter
\providecommand \@ifxundefined [1]{%
 \@ifx{#1\undefined}
}%
\providecommand \@ifnum [1]{%
 \ifnum #1\expandafter \@firstoftwo
 \else \expandafter \@secondoftwo
 \fi
}%
\providecommand \@ifx [1]{%
 \ifx #1\expandafter \@firstoftwo
 \else \expandafter \@secondoftwo
 \fi
}%
\providecommand \natexlab [1]{#1}%
\providecommand \enquote  [1]{``#1''}%
\providecommand \bibnamefont  [1]{#1}%
\providecommand \bibfnamefont [1]{#1}%
\providecommand \citenamefont [1]{#1}%
\providecommand \href@noop [0]{\@secondoftwo}%
\providecommand \href [0]{\begingroup \@sanitize@url \@href}%
\providecommand \@href[1]{\@@startlink{#1}\@@href}%
\providecommand \@@href[1]{\endgroup#1\@@endlink}%
\providecommand \@sanitize@url [0]{\catcode `\\12\catcode `\$12\catcode
  `\&12\catcode `\#12\catcode `\^12\catcode `\_12\catcode `\%12\relax}%
\providecommand \@@startlink[1]{}%
\providecommand \@@endlink[0]{}%
\providecommand \url  [0]{\begingroup\@sanitize@url \@url }%
\providecommand \@url [1]{\endgroup\@href {#1}{\urlprefix }}%
\providecommand \urlprefix  [0]{URL }%
\providecommand \Eprint [0]{\href }%
\providecommand \doibase [0]{https://doi.org/}%
\providecommand \selectlanguage [0]{\@gobble}%
\providecommand \bibinfo  [0]{\@secondoftwo}%
\providecommand \bibfield  [0]{\@secondoftwo}%
\providecommand \translation [1]{[#1]}%
\providecommand \BibitemOpen [0]{}%
\providecommand \bibitemStop [0]{}%
\providecommand \bibitemNoStop [0]{.\EOS\space}%
\providecommand \EOS [0]{\spacefactor3000\relax}%
\providecommand \BibitemShut  [1]{\csname bibitem#1\endcsname}%
\let\auto@bib@innerbib\@empty
\bibitem [{\citenamefont {Orlandini}\ and\ \citenamefont
  {Micheletti}(2021)}]{orlandini2021topological}%
  \BibitemOpen
  \bibfield  {author} {\bibinfo {author} {\bibfnamefont {E.}~\bibnamefont
  {Orlandini}}\ and\ \bibinfo {author} {\bibfnamefont {C.}~\bibnamefont
  {Micheletti}},\ }\href@noop {} {\bibfield  {journal} {\bibinfo  {journal}
  {Journal of Physics: Condensed Matter}\ }\textbf {\bibinfo {volume} {34}},\
  \bibinfo {pages} {013002} (\bibinfo {year} {2021})}\BibitemShut {NoStop}%
\bibitem [{\citenamefont {Lewis}\ \emph {et~al.}(2017)\citenamefont {Lewis},
  \citenamefont {Galli},\ and\ \citenamefont {Goldup}}]{lewis2017properties}%
  \BibitemOpen
  \bibfield  {author} {\bibinfo {author} {\bibfnamefont {J.~E.}\ \bibnamefont
  {Lewis}}, \bibinfo {author} {\bibfnamefont {M.}~\bibnamefont {Galli}},\ and\
  \bibinfo {author} {\bibfnamefont {S.~M.}\ \bibnamefont {Goldup}},\
  }\href@noop {} {\bibfield  {journal} {\bibinfo  {journal} {Chemical
  Communications}\ }\textbf {\bibinfo {volume} {53}},\ \bibinfo {pages} {298}
  (\bibinfo {year} {2017})}\BibitemShut {NoStop}%
\bibitem [{\citenamefont {Liang}\ \emph {et~al.}(2020)\citenamefont {Liang},
  \citenamefont {Li}, \citenamefont {Tang}, \citenamefont {Komiyama},\ and\
  \citenamefont {Ariga}}]{liang2020dynamism}%
  \BibitemOpen
  \bibfield  {author} {\bibinfo {author} {\bibfnamefont {X.}~\bibnamefont
  {Liang}}, \bibinfo {author} {\bibfnamefont {L.}~\bibnamefont {Li}}, \bibinfo
  {author} {\bibfnamefont {J.}~\bibnamefont {Tang}}, \bibinfo {author}
  {\bibfnamefont {M.}~\bibnamefont {Komiyama}},\ and\ \bibinfo {author}
  {\bibfnamefont {K.}~\bibnamefont {Ariga}},\ }\href@noop {} {\bibfield
  {journal} {\bibinfo  {journal} {Bulletin of the Chemical Society of Japan}\
  }\textbf {\bibinfo {volume} {93}},\ \bibinfo {pages} {581} (\bibinfo {year}
  {2020})}\BibitemShut {NoStop}%
\bibitem [{\citenamefont {Kapnistos}\ \emph {et~al.}(2008)\citenamefont
  {Kapnistos}, \citenamefont {Lang}, \citenamefont {Vlassopoulos},
  \citenamefont {Pyckhout-Hintzen}, \citenamefont {Richter}, \citenamefont
  {Cho}, \citenamefont {Chang},\ and\ \citenamefont
  {Rubinstein}}]{kapnistos2008unexpected}%
  \BibitemOpen
  \bibfield  {author} {\bibinfo {author} {\bibfnamefont {M.}~\bibnamefont
  {Kapnistos}}, \bibinfo {author} {\bibfnamefont {M.}~\bibnamefont {Lang}},
  \bibinfo {author} {\bibfnamefont {D.}~\bibnamefont {Vlassopoulos}}, \bibinfo
  {author} {\bibfnamefont {W.}~\bibnamefont {Pyckhout-Hintzen}}, \bibinfo
  {author} {\bibfnamefont {D.}~\bibnamefont {Richter}}, \bibinfo {author}
  {\bibfnamefont {D.}~\bibnamefont {Cho}}, \bibinfo {author} {\bibfnamefont
  {T.}~\bibnamefont {Chang}},\ and\ \bibinfo {author} {\bibfnamefont
  {M.}~\bibnamefont {Rubinstein}},\ }\href@noop {} {\bibfield  {journal}
  {\bibinfo  {journal} {Nature materials}\ }\textbf {\bibinfo {volume} {7}},\
  \bibinfo {pages} {997} (\bibinfo {year} {2008})}\BibitemShut {NoStop}%
\bibitem [{\citenamefont {De~Gennes}\ and\ \citenamefont
  {Gennes}(1979)}]{de1979scaling}%
  \BibitemOpen
  \bibfield  {author} {\bibinfo {author} {\bibfnamefont {P.-G.}\ \bibnamefont
  {De~Gennes}}\ and\ \bibinfo {author} {\bibfnamefont {P.-G.}\ \bibnamefont
  {Gennes}},\ }\href@noop {} {\emph {\bibinfo {title} {Scaling concepts in
  polymer physics}}}\ (\bibinfo  {publisher} {Cornell university press},\
  \bibinfo {year} {1979})\BibitemShut {NoStop}%
\bibitem [{\citenamefont {Fischer}\ \emph {et~al.}(2015)\citenamefont
  {Fischer}, \citenamefont {Lang},\ and\ \citenamefont
  {Sommer}}]{fischer2015formation}%
  \BibitemOpen
  \bibfield  {author} {\bibinfo {author} {\bibfnamefont {J.}~\bibnamefont
  {Fischer}}, \bibinfo {author} {\bibfnamefont {M.}~\bibnamefont {Lang}},\ and\
  \bibinfo {author} {\bibfnamefont {J.-U.}\ \bibnamefont {Sommer}},\
  }\href@noop {} {\bibfield  {journal} {\bibinfo  {journal} {The Journal of
  chemical physics}\ }\textbf {\bibinfo {volume} {143}},\ \bibinfo {pages}
  {243114} (\bibinfo {year} {2015})}\BibitemShut {NoStop}%
\bibitem [{\citenamefont {Krajina}\ \emph {et~al.}(2018)\citenamefont
  {Krajina}, \citenamefont {Zhu}, \citenamefont {Heilshorn},\ and\
  \citenamefont {Spakowitz}}]{krajina2018active}%
  \BibitemOpen
  \bibfield  {author} {\bibinfo {author} {\bibfnamefont {B.~A.}\ \bibnamefont
  {Krajina}}, \bibinfo {author} {\bibfnamefont {A.}~\bibnamefont {Zhu}},
  \bibinfo {author} {\bibfnamefont {S.~C.}\ \bibnamefont {Heilshorn}},\ and\
  \bibinfo {author} {\bibfnamefont {A.~J.}\ \bibnamefont {Spakowitz}},\
  }\href@noop {} {\bibfield  {journal} {\bibinfo  {journal} {Physical review
  letters}\ }\textbf {\bibinfo {volume} {121}},\ \bibinfo {pages} {148001}
  (\bibinfo {year} {2018})}\BibitemShut {NoStop}%
\bibitem [{\citenamefont {Chen}\ \emph {et~al.}(1995)\citenamefont {Chen},
  \citenamefont {Rauch}, \citenamefont {White}, \citenamefont {Englund},\ and\
  \citenamefont {Cozzarelli}}]{chen1995topology}%
  \BibitemOpen
  \bibfield  {author} {\bibinfo {author} {\bibfnamefont {J.}~\bibnamefont
  {Chen}}, \bibinfo {author} {\bibfnamefont {C.~A.}\ \bibnamefont {Rauch}},
  \bibinfo {author} {\bibfnamefont {J.~H.}\ \bibnamefont {White}}, \bibinfo
  {author} {\bibfnamefont {P.~T.}\ \bibnamefont {Englund}},\ and\ \bibinfo
  {author} {\bibfnamefont {N.~R.}\ \bibnamefont {Cozzarelli}},\ }\href@noop {}
  {\bibfield  {journal} {\bibinfo  {journal} {Cell}\ }\textbf {\bibinfo
  {volume} {80}},\ \bibinfo {pages} {61} (\bibinfo {year} {1995})}\BibitemShut
  {NoStop}%
\bibitem [{\citenamefont {Klotz}\ \emph {et~al.}(2020)\citenamefont {Klotz},
  \citenamefont {Soh},\ and\ \citenamefont {Doyle}}]{klotz2020equilibrium}%
  \BibitemOpen
  \bibfield  {author} {\bibinfo {author} {\bibfnamefont {A.~R.}\ \bibnamefont
  {Klotz}}, \bibinfo {author} {\bibfnamefont {B.~W.}\ \bibnamefont {Soh}},\
  and\ \bibinfo {author} {\bibfnamefont {P.~S.}\ \bibnamefont {Doyle}},\
  }\href@noop {} {\bibfield  {journal} {\bibinfo  {journal} {Proceedings of the
  National Academy of Sciences}\ }\textbf {\bibinfo {volume} {117}},\ \bibinfo
  {pages} {121} (\bibinfo {year} {2020})}\BibitemShut {NoStop}%
\bibitem [{\citenamefont {Soh}\ and\ \citenamefont
  {Doyle}(2020)}]{soh2020deformation}%
  \BibitemOpen
  \bibfield  {author} {\bibinfo {author} {\bibfnamefont {B.~W.}\ \bibnamefont
  {Soh}}\ and\ \bibinfo {author} {\bibfnamefont {P.~S.}\ \bibnamefont
  {Doyle}},\ }\href@noop {} {\bibfield  {journal} {\bibinfo  {journal} {ACS
  Macro Letters}\ }\textbf {\bibinfo {volume} {9}},\ \bibinfo {pages} {944}
  (\bibinfo {year} {2020})}\BibitemShut {NoStop}%
\bibitem [{\citenamefont {Polson}\ \emph {et~al.}(2021)\citenamefont {Polson},
  \citenamefont {Garcia},\ and\ \citenamefont {Klotz}}]{polson2021flatness}%
  \BibitemOpen
  \bibfield  {author} {\bibinfo {author} {\bibfnamefont {J.~M.}\ \bibnamefont
  {Polson}}, \bibinfo {author} {\bibfnamefont {E.~J.}\ \bibnamefont {Garcia}},\
  and\ \bibinfo {author} {\bibfnamefont {A.~R.}\ \bibnamefont {Klotz}},\
  }\href@noop {} {\bibfield  {journal} {\bibinfo  {journal} {Soft Matter}\
  }\textbf {\bibinfo {volume} {17}},\ \bibinfo {pages} {10505} (\bibinfo {year}
  {2021})}\BibitemShut {NoStop}%
\bibitem [{\citenamefont {Chen}\ \emph {et~al.}(2011)\citenamefont {Chen},
  \citenamefont {Bae},\ and\ \citenamefont {Granick}}]{chen2011directed}%
  \BibitemOpen
  \bibfield  {author} {\bibinfo {author} {\bibfnamefont {Q.}~\bibnamefont
  {Chen}}, \bibinfo {author} {\bibfnamefont {S.~C.}\ \bibnamefont {Bae}},\ and\
  \bibinfo {author} {\bibfnamefont {S.}~\bibnamefont {Granick}},\ }\href@noop
  {} {\bibfield  {journal} {\bibinfo  {journal} {Nature}\ }\textbf {\bibinfo
  {volume} {469}},\ \bibinfo {pages} {381} (\bibinfo {year}
  {2011})}\BibitemShut {NoStop}%
\bibitem [{\citenamefont {Wu}\ \emph {et~al.}(2017)\citenamefont {Wu},
  \citenamefont {Rauscher}, \citenamefont {Lang}, \citenamefont {Wojtecki},
  \citenamefont {De~Pablo}, \citenamefont {Hore},\ and\ \citenamefont
  {Rowan}}]{wu2017poly}%
  \BibitemOpen
  \bibfield  {author} {\bibinfo {author} {\bibfnamefont {Q.}~\bibnamefont
  {Wu}}, \bibinfo {author} {\bibfnamefont {P.~M.}\ \bibnamefont {Rauscher}},
  \bibinfo {author} {\bibfnamefont {X.}~\bibnamefont {Lang}}, \bibinfo {author}
  {\bibfnamefont {R.~J.}\ \bibnamefont {Wojtecki}}, \bibinfo {author}
  {\bibfnamefont {J.~J.}\ \bibnamefont {De~Pablo}}, \bibinfo {author}
  {\bibfnamefont {M.~J.}\ \bibnamefont {Hore}},\ and\ \bibinfo {author}
  {\bibfnamefont {S.~J.}\ \bibnamefont {Rowan}},\ }\href@noop {} {\bibfield
  {journal} {\bibinfo  {journal} {Science}\ }\textbf {\bibinfo {volume}
  {358}},\ \bibinfo {pages} {1434} (\bibinfo {year} {2017})}\BibitemShut
  {NoStop}%
\bibitem [{\citenamefont {Datta}\ \emph {et~al.}(2020)\citenamefont {Datta},
  \citenamefont {Kato}, \citenamefont {Higashiharaguchi}, \citenamefont
  {Aratsu}, \citenamefont {Isobe}, \citenamefont {Saito}, \citenamefont
  {Prabhu}, \citenamefont {Kitamoto}, \citenamefont {Hollamby}, \citenamefont
  {Smith} \emph {et~al.}}]{datta2020self}%
  \BibitemOpen
  \bibfield  {author} {\bibinfo {author} {\bibfnamefont {S.}~\bibnamefont
  {Datta}}, \bibinfo {author} {\bibfnamefont {Y.}~\bibnamefont {Kato}},
  \bibinfo {author} {\bibfnamefont {S.}~\bibnamefont {Higashiharaguchi}},
  \bibinfo {author} {\bibfnamefont {K.}~\bibnamefont {Aratsu}}, \bibinfo
  {author} {\bibfnamefont {A.}~\bibnamefont {Isobe}}, \bibinfo {author}
  {\bibfnamefont {T.}~\bibnamefont {Saito}}, \bibinfo {author} {\bibfnamefont
  {D.~D.}\ \bibnamefont {Prabhu}}, \bibinfo {author} {\bibfnamefont
  {Y.}~\bibnamefont {Kitamoto}}, \bibinfo {author} {\bibfnamefont {M.~J.}\
  \bibnamefont {Hollamby}}, \bibinfo {author} {\bibfnamefont {A.~J.}\
  \bibnamefont {Smith}}, \emph {et~al.},\ }\href@noop {} {\bibfield  {journal}
  {\bibinfo  {journal} {Nature}\ }\textbf {\bibinfo {volume} {583}},\ \bibinfo
  {pages} {400} (\bibinfo {year} {2020})}\BibitemShut {NoStop}%
\bibitem [{\citenamefont {Marenda}\ \emph {et~al.}(2018)\citenamefont
  {Marenda}, \citenamefont {Orlandini},\ and\ \citenamefont
  {Micheletti}}]{marenda2018discovering}%
  \BibitemOpen
  \bibfield  {author} {\bibinfo {author} {\bibfnamefont {M.}~\bibnamefont
  {Marenda}}, \bibinfo {author} {\bibfnamefont {E.}~\bibnamefont {Orlandini}},\
  and\ \bibinfo {author} {\bibfnamefont {C.}~\bibnamefont {Micheletti}},\
  }\href@noop {} {\bibfield  {journal} {\bibinfo  {journal} {Nature
  communications}\ }\textbf {\bibinfo {volume} {9}},\ \bibinfo {pages} {1}
  (\bibinfo {year} {2018})}\BibitemShut {NoStop}%
\bibitem [{\citenamefont {Ubertini}\ and\ \citenamefont
  {Rosa}(2021)}]{ubertini2021computer}%
  \BibitemOpen
  \bibfield  {author} {\bibinfo {author} {\bibfnamefont {M.~A.}\ \bibnamefont
  {Ubertini}}\ and\ \bibinfo {author} {\bibfnamefont {A.}~\bibnamefont
  {Rosa}},\ }\href@noop {} {\bibfield  {journal} {\bibinfo  {journal} {Physical
  Review E}\ }\textbf {\bibinfo {volume} {104}},\ \bibinfo {pages} {054503}
  (\bibinfo {year} {2021})}\BibitemShut {NoStop}%
\bibitem [{\citenamefont {Livoreil}\ \emph {et~al.}(1994)\citenamefont
  {Livoreil}, \citenamefont {Dietrich-Buchecker},\ and\ \citenamefont
  {Sauvage}}]{livoreil1994electrochemically}%
  \BibitemOpen
  \bibfield  {author} {\bibinfo {author} {\bibfnamefont {A.}~\bibnamefont
  {Livoreil}}, \bibinfo {author} {\bibfnamefont {C.~O.}\ \bibnamefont
  {Dietrich-Buchecker}},\ and\ \bibinfo {author} {\bibfnamefont {J.-P.}\
  \bibnamefont {Sauvage}},\ }\href@noop {} {\bibfield  {journal} {\bibinfo
  {journal} {Journal of the American Chemical Society}\ }\textbf {\bibinfo
  {volume} {116}},\ \bibinfo {pages} {9399} (\bibinfo {year}
  {1994})}\BibitemShut {NoStop}%
\bibitem [{\citenamefont {Hu}\ \emph {et~al.}(2015)\citenamefont {Hu},
  \citenamefont {Lu},\ and\ \citenamefont {Willner}}]{hu2015switchable}%
  \BibitemOpen
  \bibfield  {author} {\bibinfo {author} {\bibfnamefont {L.}~\bibnamefont
  {Hu}}, \bibinfo {author} {\bibfnamefont {C.-H.}\ \bibnamefont {Lu}},\ and\
  \bibinfo {author} {\bibfnamefont {I.}~\bibnamefont {Willner}},\ }\href@noop
  {} {\bibfield  {journal} {\bibinfo  {journal} {Nano letters}\ }\textbf
  {\bibinfo {volume} {15}},\ \bibinfo {pages} {2099} (\bibinfo {year}
  {2015})}\BibitemShut {NoStop}%
\bibitem [{\citenamefont {Aucagne}\ \emph {et~al.}(2007)\citenamefont
  {Aucagne}, \citenamefont {Bern{\'a}}, \citenamefont {Crowley}, \citenamefont
  {Goldup}, \citenamefont {H{\"a}nni}, \citenamefont {Leigh}, \citenamefont
  {Lusby}, \citenamefont {Ronaldson}, \citenamefont {Slawin}, \citenamefont
  {Viterisi} \emph {et~al.}}]{aucagne2007catalytic}%
  \BibitemOpen
  \bibfield  {author} {\bibinfo {author} {\bibfnamefont {V.}~\bibnamefont
  {Aucagne}}, \bibinfo {author} {\bibfnamefont {J.}~\bibnamefont {Bern{\'a}}},
  \bibinfo {author} {\bibfnamefont {J.~D.}\ \bibnamefont {Crowley}}, \bibinfo
  {author} {\bibfnamefont {S.~M.}\ \bibnamefont {Goldup}}, \bibinfo {author}
  {\bibfnamefont {K.~D.}\ \bibnamefont {H{\"a}nni}}, \bibinfo {author}
  {\bibfnamefont {D.~A.}\ \bibnamefont {Leigh}}, \bibinfo {author}
  {\bibfnamefont {P.~J.}\ \bibnamefont {Lusby}}, \bibinfo {author}
  {\bibfnamefont {V.~E.}\ \bibnamefont {Ronaldson}}, \bibinfo {author}
  {\bibfnamefont {A.~M.}\ \bibnamefont {Slawin}}, \bibinfo {author}
  {\bibfnamefont {A.}~\bibnamefont {Viterisi}}, \emph {et~al.},\ }\href@noop {}
  {\bibfield  {journal} {\bibinfo  {journal} {Journal of the American Chemical
  Society}\ }\textbf {\bibinfo {volume} {129}},\ \bibinfo {pages} {11950}
  (\bibinfo {year} {2007})}\BibitemShut {NoStop}%
\bibitem [{\citenamefont {Ayme}\ \emph {et~al.}(2015)\citenamefont {Ayme},
  \citenamefont {Beves}, \citenamefont {Campbell}, \citenamefont
  {Gil-Ram{\'\i}rez}, \citenamefont {Leigh},\ and\ \citenamefont
  {Stephens}}]{ayme2015strong}%
  \BibitemOpen
  \bibfield  {author} {\bibinfo {author} {\bibfnamefont {J.-F.}\ \bibnamefont
  {Ayme}}, \bibinfo {author} {\bibfnamefont {J.~E.}\ \bibnamefont {Beves}},
  \bibinfo {author} {\bibfnamefont {C.~J.}\ \bibnamefont {Campbell}}, \bibinfo
  {author} {\bibfnamefont {G.}~\bibnamefont {Gil-Ram{\'\i}rez}}, \bibinfo
  {author} {\bibfnamefont {D.~A.}\ \bibnamefont {Leigh}},\ and\ \bibinfo
  {author} {\bibfnamefont {A.~J.}\ \bibnamefont {Stephens}},\ }\href@noop {}
  {\bibfield  {journal} {\bibinfo  {journal} {Journal of the American Chemical
  Society}\ }\textbf {\bibinfo {volume} {137}},\ \bibinfo {pages} {9812}
  (\bibinfo {year} {2015})}\BibitemShut {NoStop}%
\bibitem [{\citenamefont {Liu}\ \emph {et~al.}(2016)\citenamefont {Liu},
  \citenamefont {Ma}, \citenamefont {Zhao}, \citenamefont {Sun}, \citenamefont
  {G{\'a}ndara}, \citenamefont {Furukawa}, \citenamefont {Liu}, \citenamefont
  {Zhu}, \citenamefont {Zhu}, \citenamefont {Suenaga} \emph
  {et~al.}}]{liu2016weaving}%
  \BibitemOpen
  \bibfield  {author} {\bibinfo {author} {\bibfnamefont {Y.}~\bibnamefont
  {Liu}}, \bibinfo {author} {\bibfnamefont {Y.}~\bibnamefont {Ma}}, \bibinfo
  {author} {\bibfnamefont {Y.}~\bibnamefont {Zhao}}, \bibinfo {author}
  {\bibfnamefont {X.}~\bibnamefont {Sun}}, \bibinfo {author} {\bibfnamefont
  {F.}~\bibnamefont {G{\'a}ndara}}, \bibinfo {author} {\bibfnamefont
  {H.}~\bibnamefont {Furukawa}}, \bibinfo {author} {\bibfnamefont
  {Z.}~\bibnamefont {Liu}}, \bibinfo {author} {\bibfnamefont {H.}~\bibnamefont
  {Zhu}}, \bibinfo {author} {\bibfnamefont {C.}~\bibnamefont {Zhu}}, \bibinfo
  {author} {\bibfnamefont {K.}~\bibnamefont {Suenaga}}, \emph {et~al.},\
  }\href@noop {} {\bibfield  {journal} {\bibinfo  {journal} {Science}\ }\textbf
  {\bibinfo {volume} {351}},\ \bibinfo {pages} {365} (\bibinfo {year}
  {2016})}\BibitemShut {NoStop}%
\bibitem [{\citenamefont {Nguyen}\ \emph {et~al.}(2005)\citenamefont {Nguyen},
  \citenamefont {Tseng}, \citenamefont {Celestre}, \citenamefont {Flood},
  \citenamefont {Liu}, \citenamefont {Stoddart},\ and\ \citenamefont
  {Zink}}]{nguyen2005reversible}%
  \BibitemOpen
  \bibfield  {author} {\bibinfo {author} {\bibfnamefont {T.~D.}\ \bibnamefont
  {Nguyen}}, \bibinfo {author} {\bibfnamefont {H.-R.}\ \bibnamefont {Tseng}},
  \bibinfo {author} {\bibfnamefont {P.~C.}\ \bibnamefont {Celestre}}, \bibinfo
  {author} {\bibfnamefont {A.~H.}\ \bibnamefont {Flood}}, \bibinfo {author}
  {\bibfnamefont {Y.}~\bibnamefont {Liu}}, \bibinfo {author} {\bibfnamefont
  {J.~F.}\ \bibnamefont {Stoddart}},\ and\ \bibinfo {author} {\bibfnamefont
  {J.~I.}\ \bibnamefont {Zink}},\ }\href@noop {} {\bibfield  {journal}
  {\bibinfo  {journal} {Proceedings of the National Academy of Sciences}\
  }\textbf {\bibinfo {volume} {102}},\ \bibinfo {pages} {10029} (\bibinfo
  {year} {2005})}\BibitemShut {NoStop}%
\bibitem [{\citenamefont {Leigh}\ \emph {et~al.}(2003)\citenamefont {Leigh},
  \citenamefont {Wong}, \citenamefont {Dehez},\ and\ \citenamefont
  {Zerbetto}}]{leigh2003unidirectional}%
  \BibitemOpen
  \bibfield  {author} {\bibinfo {author} {\bibfnamefont {D.~A.}\ \bibnamefont
  {Leigh}}, \bibinfo {author} {\bibfnamefont {J.~K.}\ \bibnamefont {Wong}},
  \bibinfo {author} {\bibfnamefont {F.}~\bibnamefont {Dehez}},\ and\ \bibinfo
  {author} {\bibfnamefont {F.}~\bibnamefont {Zerbetto}},\ }\href@noop {}
  {\bibfield  {journal} {\bibinfo  {journal} {Nature}\ }\textbf {\bibinfo
  {volume} {424}},\ \bibinfo {pages} {174} (\bibinfo {year}
  {2003})}\BibitemShut {NoStop}%
\bibitem [{\citenamefont {Hern{\'a}ndez}\ \emph {et~al.}(2004)\citenamefont
  {Hern{\'a}ndez}, \citenamefont {Kay},\ and\ \citenamefont
  {Leigh}}]{hernandez2004reversible}%
  \BibitemOpen
  \bibfield  {author} {\bibinfo {author} {\bibfnamefont {J.~V.}\ \bibnamefont
  {Hern{\'a}ndez}}, \bibinfo {author} {\bibfnamefont {E.~R.}\ \bibnamefont
  {Kay}},\ and\ \bibinfo {author} {\bibfnamefont {D.~A.}\ \bibnamefont
  {Leigh}},\ }\href@noop {} {\bibfield  {journal} {\bibinfo  {journal}
  {Science}\ }\textbf {\bibinfo {volume} {306}},\ \bibinfo {pages} {1532}
  (\bibinfo {year} {2004})}\BibitemShut {NoStop}%
\bibitem [{\citenamefont {Sauvage}\ \emph {et~al.}(2010)\citenamefont
  {Sauvage}, \citenamefont {Collin}, \citenamefont {Durot}, \citenamefont
  {Frey}, \citenamefont {Heitz}, \citenamefont {Sour},\ and\ \citenamefont
  {Tock}}]{SAUVAGE2010315}%
  \BibitemOpen
  \bibfield  {author} {\bibinfo {author} {\bibfnamefont {J.-P.}\ \bibnamefont
  {Sauvage}}, \bibinfo {author} {\bibfnamefont {J.-P.}\ \bibnamefont {Collin}},
  \bibinfo {author} {\bibfnamefont {S.}~\bibnamefont {Durot}}, \bibinfo
  {author} {\bibfnamefont {J.}~\bibnamefont {Frey}}, \bibinfo {author}
  {\bibfnamefont {V.}~\bibnamefont {Heitz}}, \bibinfo {author} {\bibfnamefont
  {A.}~\bibnamefont {Sour}},\ and\ \bibinfo {author} {\bibfnamefont
  {C.}~\bibnamefont {Tock}},\ }\href@noop {} {\bibfield  {journal} {\bibinfo
  {journal} {Comptes Rendus Chimie}\ }\textbf {\bibinfo {volume} {13}},\
  \bibinfo {pages} {315} (\bibinfo {year} {2010})}\BibitemShut {NoStop}%
\bibitem [{\citenamefont {Erbas-Cakmak}\ \emph {et~al.}(2015)\citenamefont
  {Erbas-Cakmak}, \citenamefont {Leigh}, \citenamefont {McTernan},\ and\
  \citenamefont {Nussbaumer}}]{erbas2015artificial}%
  \BibitemOpen
  \bibfield  {author} {\bibinfo {author} {\bibfnamefont {S.}~\bibnamefont
  {Erbas-Cakmak}}, \bibinfo {author} {\bibfnamefont {D.~A.}\ \bibnamefont
  {Leigh}}, \bibinfo {author} {\bibfnamefont {C.~T.}\ \bibnamefont
  {McTernan}},\ and\ \bibinfo {author} {\bibfnamefont {A.~L.}\ \bibnamefont
  {Nussbaumer}},\ }\href@noop {} {\bibfield  {journal} {\bibinfo  {journal}
  {Chemical reviews}\ }\textbf {\bibinfo {volume} {115}},\ \bibinfo {pages}
  {10081} (\bibinfo {year} {2015})}\BibitemShut {NoStop}%
\bibitem [{\citenamefont {Sauvage}(2017)}]{sauvage2017chemical}%
  \BibitemOpen
  \bibfield  {author} {\bibinfo {author} {\bibfnamefont {J.-P.}\ \bibnamefont
  {Sauvage}},\ }\href@noop {} {\bibfield  {journal} {\bibinfo  {journal}
  {Angewandte Chemie International Edition}\ }\textbf {\bibinfo {volume}
  {56}},\ \bibinfo {pages} {11080} (\bibinfo {year} {2017})}\BibitemShut
  {NoStop}%
\bibitem [{\citenamefont {Mena-Hernando}\ and\ \citenamefont
  {P{\'e}rez}(2019)}]{mena2019mechanically}%
  \BibitemOpen
  \bibfield  {author} {\bibinfo {author} {\bibfnamefont {S.}~\bibnamefont
  {Mena-Hernando}}\ and\ \bibinfo {author} {\bibfnamefont {E.~M.}\ \bibnamefont
  {P{\'e}rez}},\ }\href@noop {} {\bibfield  {journal} {\bibinfo  {journal}
  {Chemical Society Reviews}\ }\textbf {\bibinfo {volume} {48}},\ \bibinfo
  {pages} {5016} (\bibinfo {year} {2019})}\BibitemShut {NoStop}%
\bibitem [{\citenamefont {Rauscher}\ \emph {et~al.}(2018)\citenamefont
  {Rauscher}, \citenamefont {Rowan},\ and\ \citenamefont
  {de~Pablo}}]{rauscher2018topological}%
  \BibitemOpen
  \bibfield  {author} {\bibinfo {author} {\bibfnamefont {P.~M.}\ \bibnamefont
  {Rauscher}}, \bibinfo {author} {\bibfnamefont {S.~J.}\ \bibnamefont
  {Rowan}},\ and\ \bibinfo {author} {\bibfnamefont {J.~J.}\ \bibnamefont
  {de~Pablo}},\ }\href@noop {} {\bibfield  {journal} {\bibinfo  {journal} {ACS
  Macro Letters}\ }\textbf {\bibinfo {volume} {7}},\ \bibinfo {pages} {938}
  (\bibinfo {year} {2018})}\BibitemShut {NoStop}%
\bibitem [{\citenamefont {Rauscher}\ \emph {et~al.}(2020)\citenamefont
  {Rauscher}, \citenamefont {Schweizer}, \citenamefont {Rowan},\ and\
  \citenamefont {de~Pablo}}]{rauscher2020dynamics}%
  \BibitemOpen
  \bibfield  {author} {\bibinfo {author} {\bibfnamefont {P.~M.}\ \bibnamefont
  {Rauscher}}, \bibinfo {author} {\bibfnamefont {K.~S.}\ \bibnamefont
  {Schweizer}}, \bibinfo {author} {\bibfnamefont {S.~J.}\ \bibnamefont
  {Rowan}},\ and\ \bibinfo {author} {\bibfnamefont {J.~J.}\ \bibnamefont
  {de~Pablo}},\ }\href@noop {} {\bibfield  {journal} {\bibinfo  {journal} {The
  Journal of Chemical Physics}\ }\textbf {\bibinfo {volume} {152}},\ \bibinfo
  {pages} {214901} (\bibinfo {year} {2020})}\BibitemShut {NoStop}%
\bibitem [{\citenamefont {Dehaghani}\ \emph {et~al.}(2020)\citenamefont
  {Dehaghani}, \citenamefont {Chubak}, \citenamefont {Likos},\ and\
  \citenamefont {Ejtehadi}}]{dehaghani2020effects}%
  \BibitemOpen
  \bibfield  {author} {\bibinfo {author} {\bibfnamefont {Z.~A.}\ \bibnamefont
  {Dehaghani}}, \bibinfo {author} {\bibfnamefont {I.}~\bibnamefont {Chubak}},
  \bibinfo {author} {\bibfnamefont {C.~N.}\ \bibnamefont {Likos}},\ and\
  \bibinfo {author} {\bibfnamefont {M.~R.}\ \bibnamefont {Ejtehadi}},\
  }\href@noop {} {\bibfield  {journal} {\bibinfo  {journal} {Soft matter}\
  }\textbf {\bibinfo {volume} {16}},\ \bibinfo {pages} {3029} (\bibinfo {year}
  {2020})}\BibitemShut {NoStop}%
\bibitem [{\citenamefont {Lei}\ \emph {et~al.}(2021)\citenamefont {Lei},
  \citenamefont {Zhang}, \citenamefont {Wang},\ and\ \citenamefont
  {Zhang}}]{lei2021dimensional}%
  \BibitemOpen
  \bibfield  {author} {\bibinfo {author} {\bibfnamefont {H.}~\bibnamefont
  {Lei}}, \bibinfo {author} {\bibfnamefont {J.}~\bibnamefont {Zhang}}, \bibinfo
  {author} {\bibfnamefont {L.}~\bibnamefont {Wang}},\ and\ \bibinfo {author}
  {\bibfnamefont {G.}~\bibnamefont {Zhang}},\ }\href@noop {} {\bibfield
  {journal} {\bibinfo  {journal} {Polymer}\ }\textbf {\bibinfo {volume}
  {212}},\ \bibinfo {pages} {123160} (\bibinfo {year} {2021})}\BibitemShut
  {NoStop}%
\bibitem [{\citenamefont {Plimpton}(1995)}]{plimpton1995lammps}%
  \BibitemOpen
  \bibfield  {author} {\bibinfo {author} {\bibfnamefont {S.}~\bibnamefont
  {Plimpton}},\ }\href@noop {} {\bibfield  {journal} {\bibinfo  {journal}
  {Journal of computational physics}\ }\textbf {\bibinfo {volume} {117}},\
  \bibinfo {pages} {1} (\bibinfo {year} {1995})}\BibitemShut {NoStop}%
\bibitem [{\citenamefont {Krajina}\ and\ \citenamefont
  {Spakowitz}(2016)}]{krajina2016large}%
  \BibitemOpen
  \bibfield  {author} {\bibinfo {author} {\bibfnamefont {B.~A.}\ \bibnamefont
  {Krajina}}\ and\ \bibinfo {author} {\bibfnamefont {A.~J.}\ \bibnamefont
  {Spakowitz}},\ }\href@noop {} {\bibfield  {journal} {\bibinfo  {journal}
  {Biophysical journal}\ }\textbf {\bibinfo {volume} {111}},\ \bibinfo {pages}
  {1339} (\bibinfo {year} {2016})}\BibitemShut {NoStop}%
\bibitem [{\citenamefont {Moffatt}\ and\ \citenamefont
  {Ricca}(1995)}]{moffatt1995helicity}%
  \BibitemOpen
  \bibfield  {author} {\bibinfo {author} {\bibfnamefont {H.~K.}\ \bibnamefont
  {Moffatt}}\ and\ \bibinfo {author} {\bibfnamefont {R.~L.}\ \bibnamefont
  {Ricca}},\ }in\ \href@noop {} {\emph {\bibinfo {booktitle} {Knots And
  Applications}}}\ (\bibinfo  {publisher} {World Scientific},\ \bibinfo {year}
  {1995})\ pp.\ \bibinfo {pages} {251--269}\BibitemShut {NoStop}%
\bibitem [{\citenamefont {White}(1969)}]{white1969self}%
  \BibitemOpen
  \bibfield  {author} {\bibinfo {author} {\bibfnamefont {J.~H.}\ \bibnamefont
  {White}},\ }\href@noop {} {\bibfield  {journal} {\bibinfo  {journal}
  {American journal of mathematics}\ }\textbf {\bibinfo {volume} {91}},\
  \bibinfo {pages} {693} (\bibinfo {year} {1969})}\BibitemShut {NoStop}%
\bibitem [{\citenamefont {Bates}\ \emph {et~al.}(2005)\citenamefont {Bates},
  \citenamefont {Maxwell} \emph {et~al.}}]{bates2005dna}%
  \BibitemOpen
  \bibfield  {author} {\bibinfo {author} {\bibfnamefont {A.~D.}\ \bibnamefont
  {Bates}}, \bibinfo {author} {\bibfnamefont {A.}~\bibnamefont {Maxwell}},
  \emph {et~al.},\ }\href@noop {} {\emph {\bibinfo {title} {DNA topology}}}\
  (\bibinfo  {publisher} {Oxford University Press, USA},\ \bibinfo {year}
  {2005})\BibitemShut {NoStop}%
\bibitem [{\citenamefont {van Rensburg}\ \emph {et~al.}(1993)\citenamefont {van
  Rensburg}, \citenamefont {Orlandini}, \citenamefont {Sumners}, \citenamefont
  {Tesi},\ and\ \citenamefont {Whittington}}]{van1993writhe}%
  \BibitemOpen
  \bibfield  {author} {\bibinfo {author} {\bibfnamefont {E.~J.}\ \bibnamefont
  {van Rensburg}}, \bibinfo {author} {\bibfnamefont {E.}~\bibnamefont
  {Orlandini}}, \bibinfo {author} {\bibfnamefont {D.}~\bibnamefont {Sumners}},
  \bibinfo {author} {\bibfnamefont {M.}~\bibnamefont {Tesi}},\ and\ \bibinfo
  {author} {\bibfnamefont {S.}~\bibnamefont {Whittington}},\ }\href@noop {}
  {\bibfield  {journal} {\bibinfo  {journal} {Journal of Physics A:
  Mathematical and General}\ }\textbf {\bibinfo {volume} {26}},\ \bibinfo
  {pages} {L981} (\bibinfo {year} {1993})}\BibitemShut {NoStop}%
\bibitem [{\citenamefont {Portillo}\ \emph {et~al.}(2011)\citenamefont
  {Portillo}, \citenamefont {Diao}, \citenamefont {Scharein}, \citenamefont
  {Arsuaga},\ and\ \citenamefont {Vazquez}}]{portillo2011mean}%
  \BibitemOpen
  \bibfield  {author} {\bibinfo {author} {\bibfnamefont {J.}~\bibnamefont
  {Portillo}}, \bibinfo {author} {\bibfnamefont {Y.}~\bibnamefont {Diao}},
  \bibinfo {author} {\bibfnamefont {R.}~\bibnamefont {Scharein}}, \bibinfo
  {author} {\bibfnamefont {J.}~\bibnamefont {Arsuaga}},\ and\ \bibinfo {author}
  {\bibfnamefont {M.}~\bibnamefont {Vazquez}},\ }\href@noop {} {\bibfield
  {journal} {\bibinfo  {journal} {Journal of Physics A: Mathematical and
  Theoretical}\ }\textbf {\bibinfo {volume} {44}},\ \bibinfo {pages} {275004}
  (\bibinfo {year} {2011})}\BibitemShut {NoStop}%
\bibitem [{\citenamefont {Panagiotou}\ \emph {et~al.}(2010)\citenamefont
  {Panagiotou}, \citenamefont {Millett},\ and\ \citenamefont
  {Lambropoulou}}]{panagiotou2010linking}%
  \BibitemOpen
  \bibfield  {author} {\bibinfo {author} {\bibfnamefont {E.}~\bibnamefont
  {Panagiotou}}, \bibinfo {author} {\bibfnamefont {K.~C.}\ \bibnamefont
  {Millett}},\ and\ \bibinfo {author} {\bibfnamefont {S.}~\bibnamefont
  {Lambropoulou}},\ }\href@noop {} {\bibfield  {journal} {\bibinfo  {journal}
  {Journal of Physics A: Mathematical and Theoretical}\ }\textbf {\bibinfo
  {volume} {43}},\ \bibinfo {pages} {045208} (\bibinfo {year}
  {2010})}\BibitemShut {NoStop}%
\bibitem [{\citenamefont {Marko}(2011)}]{marko2011scaling}%
  \BibitemOpen
  \bibfield  {author} {\bibinfo {author} {\bibfnamefont {J.~F.}\ \bibnamefont
  {Marko}},\ }\href@noop {} {\bibfield  {journal} {\bibinfo  {journal} {Journal
  of statistical physics}\ }\textbf {\bibinfo {volume} {142}},\ \bibinfo
  {pages} {1353} (\bibinfo {year} {2011})}\BibitemShut {NoStop}%
\bibitem [{\citenamefont {Micheletti}\ \emph {et~al.}(2006)\citenamefont
  {Micheletti}, \citenamefont {Marenduzzo}, \citenamefont {Orlandini},\ and\
  \citenamefont {Summers}}]{micheletti2006knotting}%
  \BibitemOpen
  \bibfield  {author} {\bibinfo {author} {\bibfnamefont {C.}~\bibnamefont
  {Micheletti}}, \bibinfo {author} {\bibfnamefont {D.}~\bibnamefont
  {Marenduzzo}}, \bibinfo {author} {\bibfnamefont {E.}~\bibnamefont
  {Orlandini}},\ and\ \bibinfo {author} {\bibfnamefont {D.}~\bibnamefont
  {Summers}},\ }\href@noop {} {\bibfield  {journal} {\bibinfo  {journal} {The
  Journal of chemical physics}\ }\textbf {\bibinfo {volume} {124}},\ \bibinfo
  {pages} {064903} (\bibinfo {year} {2006})}\BibitemShut {NoStop}%
\bibitem [{\citenamefont {Baiesi}\ \emph {et~al.}(2009)\citenamefont {Baiesi},
  \citenamefont {Orlandini},\ and\ \citenamefont
  {Whittington}}]{baiesi2009interplay}%
  \BibitemOpen
  \bibfield  {author} {\bibinfo {author} {\bibfnamefont {M.}~\bibnamefont
  {Baiesi}}, \bibinfo {author} {\bibfnamefont {E.}~\bibnamefont {Orlandini}},\
  and\ \bibinfo {author} {\bibfnamefont {S.~G.}\ \bibnamefont {Whittington}},\
  }\href@noop {} {\bibfield  {journal} {\bibinfo  {journal} {The Journal of
  chemical physics}\ }\textbf {\bibinfo {volume} {131}},\ \bibinfo {pages}
  {154902} (\bibinfo {year} {2009})}\BibitemShut {NoStop}%
\bibitem [{\citenamefont {Fuller}(1978)}]{fuller1978decomposition}%
  \BibitemOpen
  \bibfield  {author} {\bibinfo {author} {\bibfnamefont {F.~B.}\ \bibnamefont
  {Fuller}},\ }\href@noop {} {\bibfield  {journal} {\bibinfo  {journal}
  {Proceedings of the National Academy of Sciences}\ }\textbf {\bibinfo
  {volume} {75}},\ \bibinfo {pages} {3557} (\bibinfo {year}
  {1978})}\BibitemShut {NoStop}%
\bibitem [{\citenamefont {Calugareanu}(1959)}]{calugareanu1959integrale}%
  \BibitemOpen
  \bibfield  {author} {\bibinfo {author} {\bibfnamefont {G.}~\bibnamefont
  {Calugareanu}},\ }\href@noop {} {\bibfield  {journal} {\bibinfo  {journal}
  {Rev. Math. pures appl}\ }\textbf {\bibinfo {volume} {4}} (\bibinfo {year}
  {1959})}\BibitemShut {NoStop}%
\bibitem [{\citenamefont {Kamien}(2002)}]{kamien2002geometry}%
  \BibitemOpen
  \bibfield  {author} {\bibinfo {author} {\bibfnamefont {R.~D.}\ \bibnamefont
  {Kamien}},\ }\href@noop {} {\bibfield  {journal} {\bibinfo  {journal}
  {Reviews of Modern physics}\ }\textbf {\bibinfo {volume} {74}},\ \bibinfo
  {pages} {953} (\bibinfo {year} {2002})}\BibitemShut {NoStop}%
\bibitem [{\citenamefont {Schmidt}\ and\ \citenamefont
  {Heckel}(2011)}]{schmidt2011construction}%
  \BibitemOpen
  \bibfield  {author} {\bibinfo {author} {\bibfnamefont {T.~L.}\ \bibnamefont
  {Schmidt}}\ and\ \bibinfo {author} {\bibfnamefont {A.}~\bibnamefont
  {Heckel}},\ }\href@noop {} {\bibfield  {journal} {\bibinfo  {journal} {Nano
  letters}\ }\textbf {\bibinfo {volume} {11}},\ \bibinfo {pages} {1739}
  (\bibinfo {year} {2011})}\BibitemShut {NoStop}%
\bibitem [{\citenamefont {Castano}\ and\ \citenamefont
  {Flatau}(2014)}]{castano2014smart}%
  \BibitemOpen
  \bibfield  {author} {\bibinfo {author} {\bibfnamefont {L.~M.}\ \bibnamefont
  {Castano}}\ and\ \bibinfo {author} {\bibfnamefont {A.~B.}\ \bibnamefont
  {Flatau}},\ }\href@noop {} {\bibfield  {journal} {\bibinfo  {journal} {Smart
  Materials and structures}\ }\textbf {\bibinfo {volume} {23}},\ \bibinfo
  {pages} {053001} (\bibinfo {year} {2014})}\BibitemShut {NoStop}%
\bibitem [{\citenamefont {Xiong}\ \emph {et~al.}(2021)\citenamefont {Xiong},
  \citenamefont {Chen},\ and\ \citenamefont {Lee}}]{xiong2021functional}%
  \BibitemOpen
  \bibfield  {author} {\bibinfo {author} {\bibfnamefont {J.}~\bibnamefont
  {Xiong}}, \bibinfo {author} {\bibfnamefont {J.}~\bibnamefont {Chen}},\ and\
  \bibinfo {author} {\bibfnamefont {P.~S.}\ \bibnamefont {Lee}},\ }\href@noop
  {} {\bibfield  {journal} {\bibinfo  {journal} {Advanced Materials}\ }\textbf
  {\bibinfo {volume} {33}},\ \bibinfo {pages} {2002640} (\bibinfo {year}
  {2021})}\BibitemShut {NoStop}%
\bibitem [{\citenamefont {Duplantier}(1981)}]{duplantier1981linking}%
  \BibitemOpen
  \bibfield  {author} {\bibinfo {author} {\bibfnamefont {B.}~\bibnamefont
  {Duplantier}},\ }\href@noop {} {\bibfield  {journal} {\bibinfo  {journal}
  {Communications in mathematical physics}\ }\textbf {\bibinfo {volume} {82}},\
  \bibinfo {pages} {41} (\bibinfo {year} {1981})}\BibitemShut {NoStop}%
\bibitem [{\citenamefont {Dennis}\ and\ \citenamefont
  {Hannay}(2005)}]{dennis2005geometry}%
  \BibitemOpen
  \bibfield  {author} {\bibinfo {author} {\bibfnamefont {M.}~\bibnamefont
  {Dennis}}\ and\ \bibinfo {author} {\bibfnamefont {J.}~\bibnamefont
  {Hannay}},\ }\href@noop {} {\bibfield  {journal} {\bibinfo  {journal}
  {Proceedings of the Royal Society A: Mathematical, Physical and Engineering
  Sciences}\ }\textbf {\bibinfo {volume} {461}},\ \bibinfo {pages} {3245}
  (\bibinfo {year} {2005})}\BibitemShut {NoStop}%
\bibitem [{\citenamefont {Klenin}\ and\ \citenamefont
  {Langowski}(2000)}]{klenin2000computation}%
  \BibitemOpen
  \bibfield  {author} {\bibinfo {author} {\bibfnamefont {K.}~\bibnamefont
  {Klenin}}\ and\ \bibinfo {author} {\bibfnamefont {J.}~\bibnamefont
  {Langowski}},\ }\href@noop {} {\bibfield  {journal} {\bibinfo  {journal}
  {Biopolymers: Original Research on Biomolecules}\ }\textbf {\bibinfo {volume}
  {54}},\ \bibinfo {pages} {307} (\bibinfo {year} {2000})}\BibitemShut
  {NoStop}%
\end{thebibliography}%
\newpage
\sifigures
\section{supplementary material}
\subsection{Polymer model}

\section{supplementary material}
\subsection{Polymer model}
Our polycatenanes are constituted by $n$ elementary rings each with $m$ beads of diameter $\sigma$.
The energy associated to each configuration is given by $H = H_{intra} + H_{inter}$, where $H_{intra}$ includes all energy terms
related to a single elementary  ring, and $H_{inter}$ is the interaction energy  between different elementary rings forming  the polycatenane.

Specifically, for each of the $n$ rings we define $H_{intra}$ as follows.
\begin{widetext}
\begin{equation}
    H_{intra} = \sum_{i=1}^{m} \left[U_{FENE}(i,i+1) + U_{bend}(i,i+1,i+2) + \\ \sum_{j=i+1}^{m}U_{WCA}(i,j)\right],
\end{equation}
\end{widetext}
where $i$ and $j$ indicate the index of the bead and the modulo on $m$ is implicitly assumed to account for the periodic nature of the rings. In the following, we use $r_{i,j}$ to indicate the norm of the vector $\mathbf{r}_i -\mathbf{r}_j$.

The FENE term, $U_{FENE}$ is defined as follows:
\begin{equation}
    U_{\scriptscriptstyle FENE}(i,i+1) =
    \begin{cases}
     -\frac{kR_0^2}{2}\ln\left[1-\left(\frac{r_{i,i+1}}{R_0}\right)^2 \right] & \text{if } r_{i,i+1} \leq R_0\\
    0 &\text{if } r_{i,i+1} > R_0.
    \end{cases}
\end{equation}
Here the maximum extension of the bond is set to $R_0 = 1.5\sigma$ and its strength to $k=30.0\epsilon/\sigma^2$.

The bending potential, $U_{bend}(i,i+1,i+2)$ is given by a Kratky-Porod term:
\begin{equation}
    U_{bend}(i,i+1,i+2) = \kappa_\theta \left(1 - \frac{\mathbf{t}_i\cdot \mathbf{t}_{i+1}}{||\mathbf{t}_i|| ||\mathbf{t}_{i+1}||}\right),
\end{equation}
where $\mathbf{t}_i = \mathbf{r}_{i+1} - \mathbf{r}_i$ is the $i-th$ bond vector, $\kappa_\theta = \frac{K_B T l_p}{\sigma}$
and $l_p$ is the persistence length. By choosing $l_p = 2m\sigma$ we consider  elementary rings that are essentially rigid. 

The steric interaction is accounted for by using the Weeks-Chandler-Andersen (WCA) potential:
\begin{equation}
    U_{WCA}(i,j) =
    \begin{cases}
     4\epsilon\left[\left(\frac{\sigma}{r_{ij}}\right)^{12} - \left(\frac{\sigma}{r_{ij}}\right)^6 \right] + \epsilon & \text{if }  r_{ij}\leq 2^{1/6}\sigma \\
    0 &\text{if } r_{ij} > 2^{1/6}\sigma \\
    \end{cases}
\end{equation}

Finally, the interaction between different rings, $H_{inter}$ is given  by:
\begin{equation}
    H_{inter} = \sum_{I\neq J}^n\sum_{i_I=1}^m\sum_{j_J=1}^m U_{WCA}(i_I, j_J),
\end{equation}
where $I$ and $J$ run over the $n$ elementary  rings while $i_I$ and $j_J$ run  over the beads belonging to ring $I$ and $J$.

\subsection{System setup}
\begin{figure*}
    \centering
    \includegraphics[width=\textwidth]{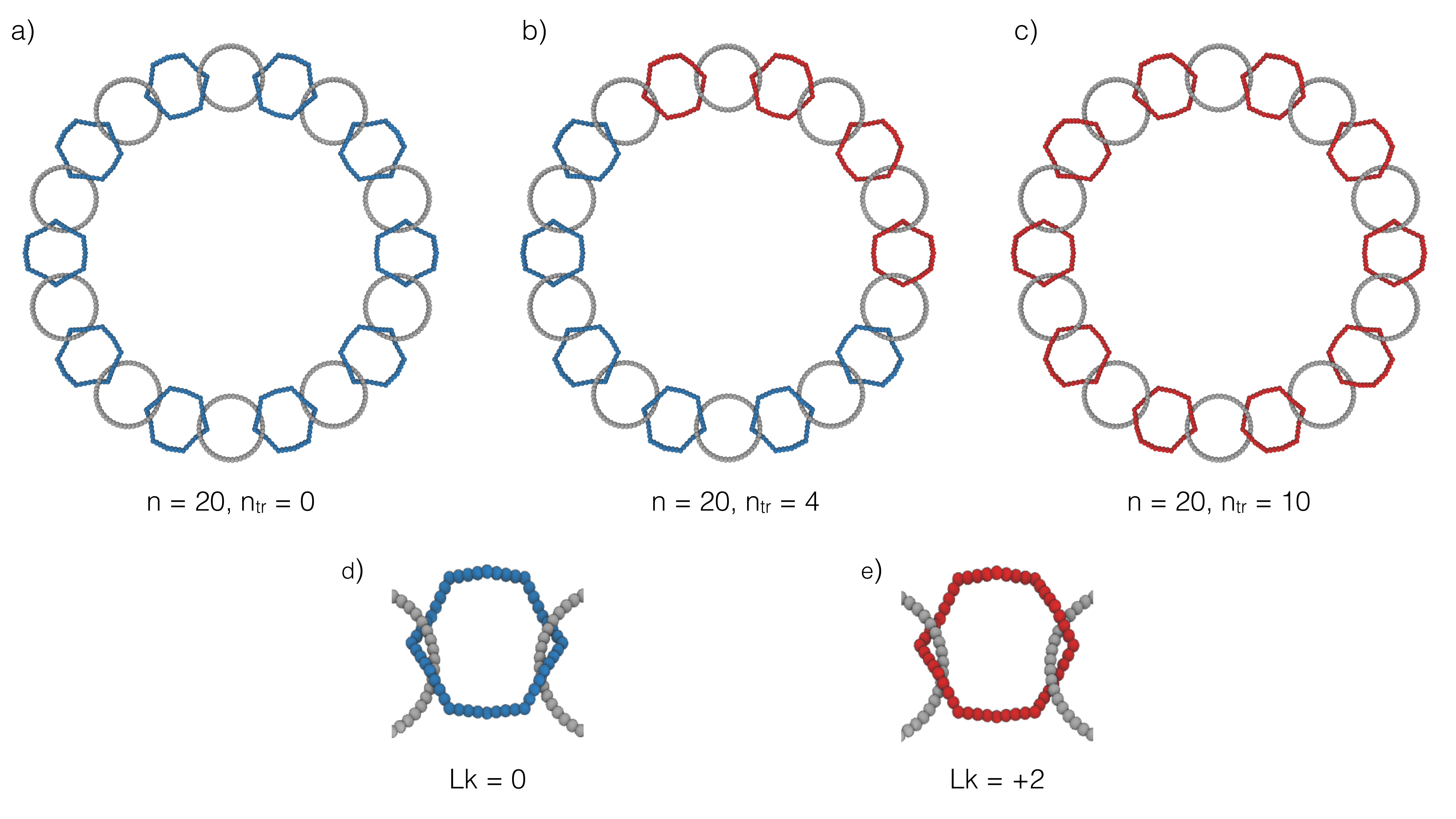}
    \caption{Initial configuration for (a) a zero torsion polycatenane, (b) a polycatenane with $n_{tr} = 4$ and (c) a polycatenane with $n_{tr} = 10$. Blue rings, (d), contribute a $+1$ and a $-1$ Hopf links, and thus have $Lk=0$; red rings, (e), contribute two $+1$ links, $Lk=2$.  }
    \label{fig:system-setup-si}
\end{figure*}
To construct an annular polycatenane with a fixed amount of quenched torsion, we proceed as follows.  Given $n=2k$ rings, we first place $k$ of them flat on even vertices of a planar $n$-gon (gray rings in Fig.~\ref{fig:system-setup-si}). Then, we complete the polycatenane by adding the rings on the odd vertices in such a way that they form either two $+1$ Hopf links with their neighbours (red rings in Fig.~\ref{fig:system-setup-si}) or a $-1$ and $+1$ Hopf link (blue rings in Fig.~\ref{fig:system-setup-si}). This is achieved by inserting suitably deformed and rotated dodecagons, as shown in Fig.~\ref{fig:system-setup-si}. While the blue dodecagons in Fig.~\ref{fig:system-setup-si} can freely rotate around the axis joining the centers of their neighbours, this is not true for the red dodecagons, which 
cannot undergo the same rotation without affecting their neighbours. This constraint due to topology introduces a torsion to the whole polycatenane. 

By varying the number $n_{tr}$ of red rings in the interval $[0,n/2]$ we can store torsion in the system from a minimum value of $0$ ($n_{tr}=0$, torsionally relaxed polycatenane) to the maximum value $\max(n_{tr})=n/2$. This corresponds to the maximum amount of twist that can be inserted into a planar circle. Higher values can be achieved by considering out-of-plane initial configurations. 

To recover the desired system of semi-rigid rings, we perform an energy minimization of the system followed by a short equilibration run of the Langevin Dynamics. This is sufficient to relax  the dodecagons in Fig.~\ref{fig:system-setup-si} into semi-rigid rings. 

\subsection{Equivalence of different circular polycatenanes with fixed $n_{tr}$}
\begin{figure}
    \includegraphics[width=1.0\columnwidth]{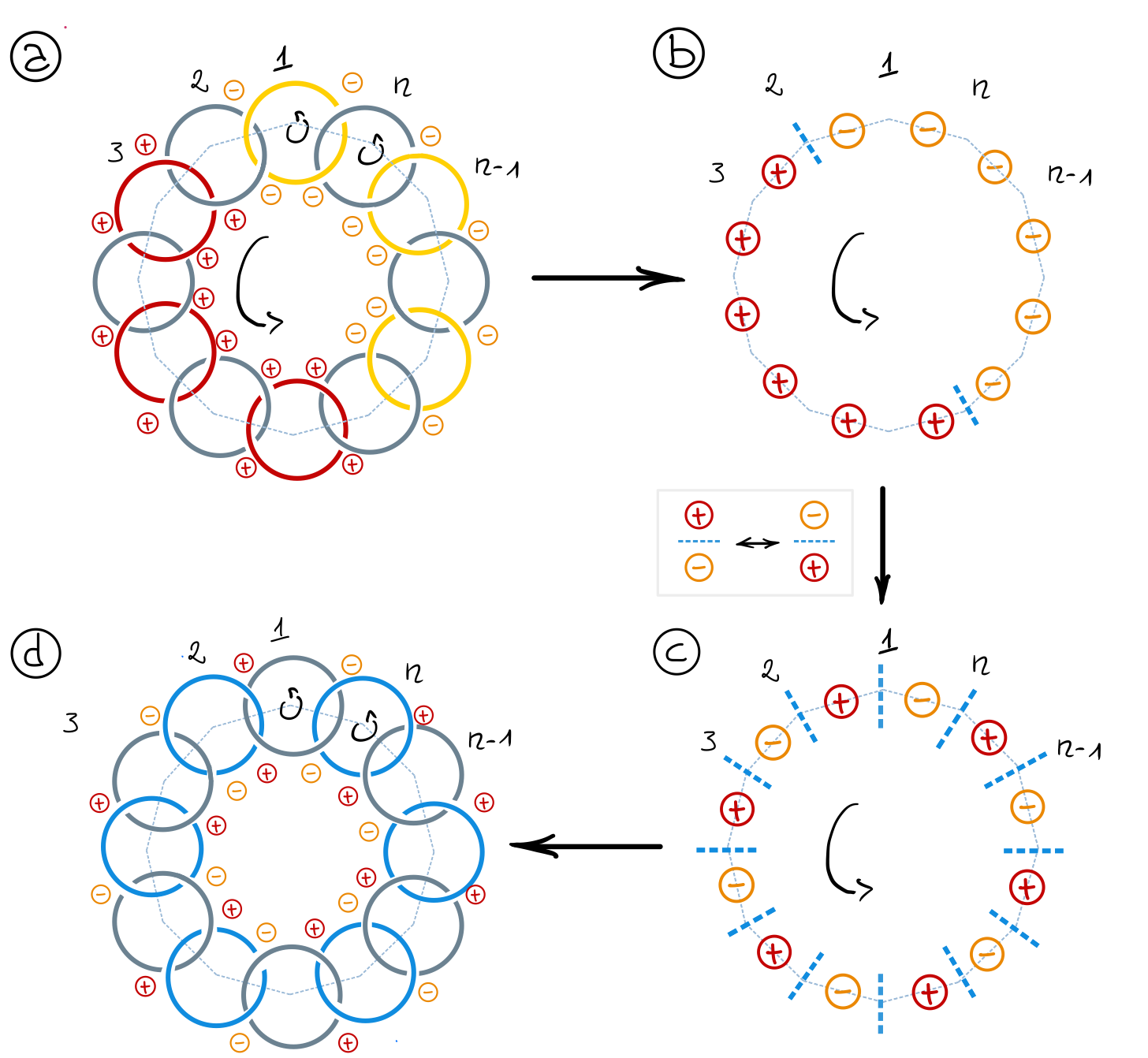}
    \caption{a) a polycatenane with $n_{tr} = 0$ obtained by using an equal number of +2 rings (red) and -2 rings (yellow). b) Its dual Ising lattice, where the spins correspond to the Hopf links. The system has magnetization $m = 2n_{tr}$. Rings that can be flipped without affecting their neighbours correspond to "walls" between a $+1$ and a $-1$ spin in the Ising system. c) A move which exchange both spins around a wall clearly preserve the magnetization $m$ and is ergodic, allowing to reach conformations such as $d$ in which there are no yellow rings.  }
    \label{fig:ergodicity}
\end{figure}

As specified above, our systems are built by inserting $n_{+} = n_{tr}$ rings forming +1 Hopf links with both neighbours ($+2$-rings), and $n/2 - n_{tr}$ rings which do not contribute to $Lk$ ($0$-rings). Note that the same amount of torsion  $n_{tr}$ can also be obtained by placing $n_{+}$ rings, each contributing to +2 Hops links ($+2$ rings) and $n_{-}$ rings each contributing to -2 Hopf links ($-2$ rings), so  that $n_{tr} = n_{+} - n_{-}$. Here we show that all polycatenanes with a given value of $n_{tr}$ have the same physical behaviour, so that the set of rings chosen  to fix $n_{tr}$ does not affect the results. To do so, we first notice that the choice of the normal of the rings is rather arbitrary, as the rings themselves do not have a physical orientation. For this reason, we can reorganize the placement of the red, blue, and yellow rings in Fig 1 in the main text by flipping a ring with a -1 and +1 Hopf links and then reassigning the normals so that they all point either up or down. To show that this move is sufficient to connect all polycatenanes with the same value of $n_{tr}$ we map our system to a circular ising chain, in which the $+1$ and $-1$ values of the spins correspond to +1 and -1 Hopf links respectively, see Fig.~\ref{fig:ergodicity}. Clearly, the ``magnetization" of this system, $m=2(n_{+}-n_{-})=2*n_{tr}$ corresponds to the injected torsion of the macroring. Note that the Ising chain can be seen as the dual of the circular  polycatenane and the elementary rings that can be flipped without affecting their neighbours correspond to domain walls separating a $+1$ and a $-1$ spin. The flipping move on the polycatenane is then equivalent to the exchange of  two spins around a wall, a move that clearly preserves the magnetization (Kawasaki move) and that can be shown to be ergodic in the space of configurations of the Ising chain with fixed $m$. Therefore, we can use only red (+2) and blue (0) rings to fix the value of $n_{tr}$ in the circular polycatenane.

\subsection{Equivalence between $n_{tr}$ and twist of a linear polycatenane}
\begin{figure}
    \includegraphics[width=1.0\columnwidth]{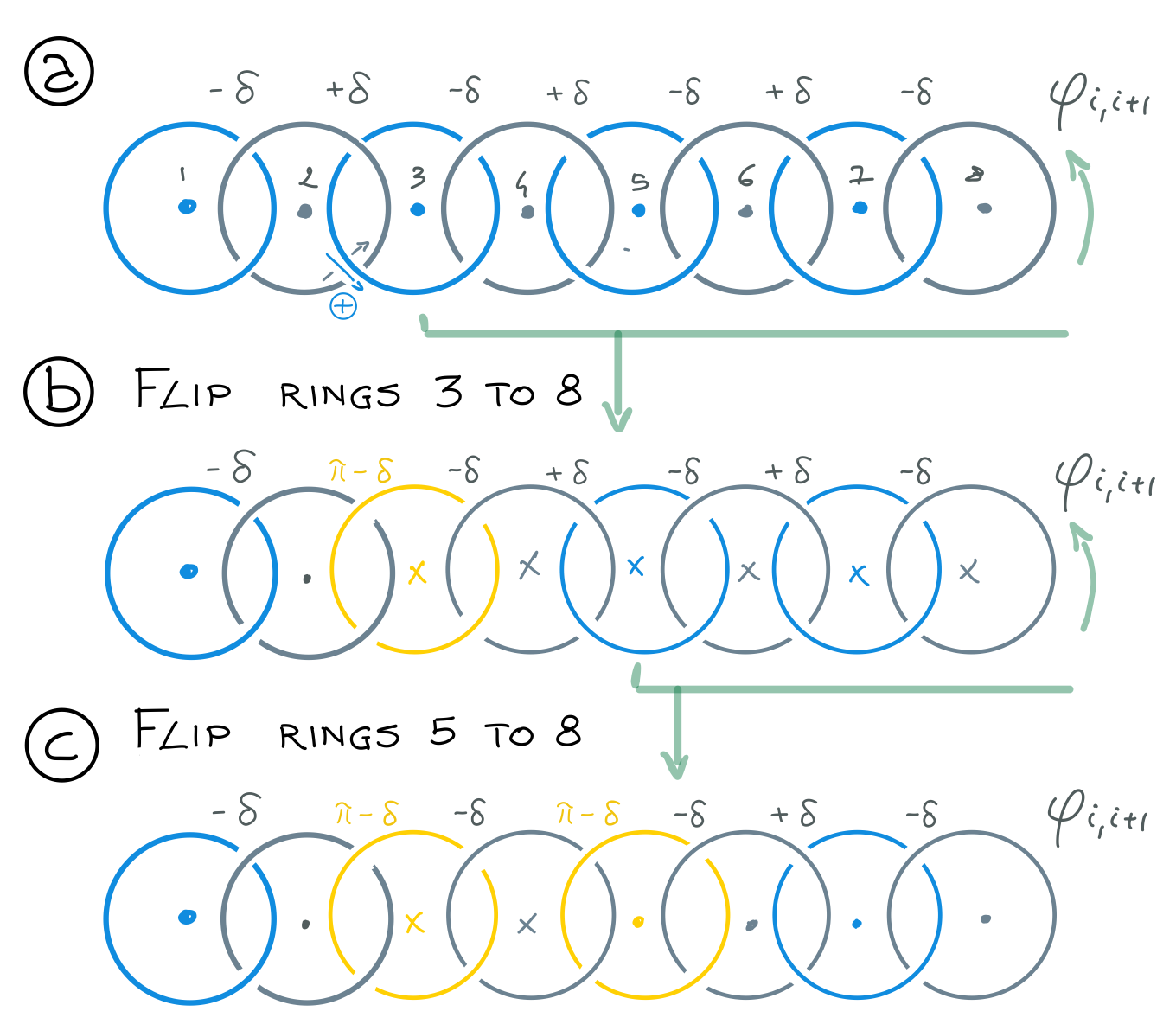}
    \caption{Passing from twist to ring-ring liking number. The greek letters above the polycatenanes indicate the twisting angle $\phi_{i,i+1}$ between rings $i$ and $i+1$. Dots and crosses in the ring centers indicate normals pointing up and down respectively. The ring colors correspond to those introduced in the main text and indicate the $Lk$ of the ring with its neighbours when all normals point up. A linear polycatenane formed by $0$-rings with their normals pointing up, a), becomes a polycatenane with $n_{tr}=2$, c), if one flips successive rings and then redefine the normals. In this simple case, each yellow ring correspond to a twist angle of $+\pi$.
    }
    \label{fig:twisting}
\end{figure}

The definition of $n_{tr}$ in our model is based on the linking number between neighboring  elementary rings, which in turn depends on their orientation. One can then notice that each insertion of a $+2$-ring and $-2$ rings corresponds to a twist of the backbone of a linear polycatenane by an angle equal to $-\pi$ and $+\pi$ respectively. This is shown in Fig.~\ref{fig:twisting}.

\subsection{Langevin Dynamics}
The  underdamped Langevin dynamics of the systems is integrated numerically  using the LAMMPS package~\cite{plimpton1995lammps} with  the  LJ units of $\sigma=M=\epsilon=K_B=1$, temperature $T=1.0$ and damping time $\tau_{damp} \equiv \frac{1}{\gamma} = 10 \tau_{LJ}$.  Starting for a given intial configuration we evolved the system for a minimum of $10^9$ steps, with timestep $\Delta t = 0.0124 \tau_{LJ}$. 

\subsection{Conformation analyses}
We characterize the equilibrium configurational properties of an annular polycatenane $\Gamma$ through observables such as the squared radius of gyration, $R^2_g(\Gamma)$, the Twist, $Tw(\Gamma)$, and the writhe $Wr(\Gamma)$ where  $\Gamma = \{\mathbf{r}_{1,1},\ldots,\mathbf{r}_{n,m}\}$ is the set of coordinates  of all beads of the elementary rings forming the polycatenane. 
\begin{figure}
    \centering
    \includegraphics[width=0.9\columnwidth]{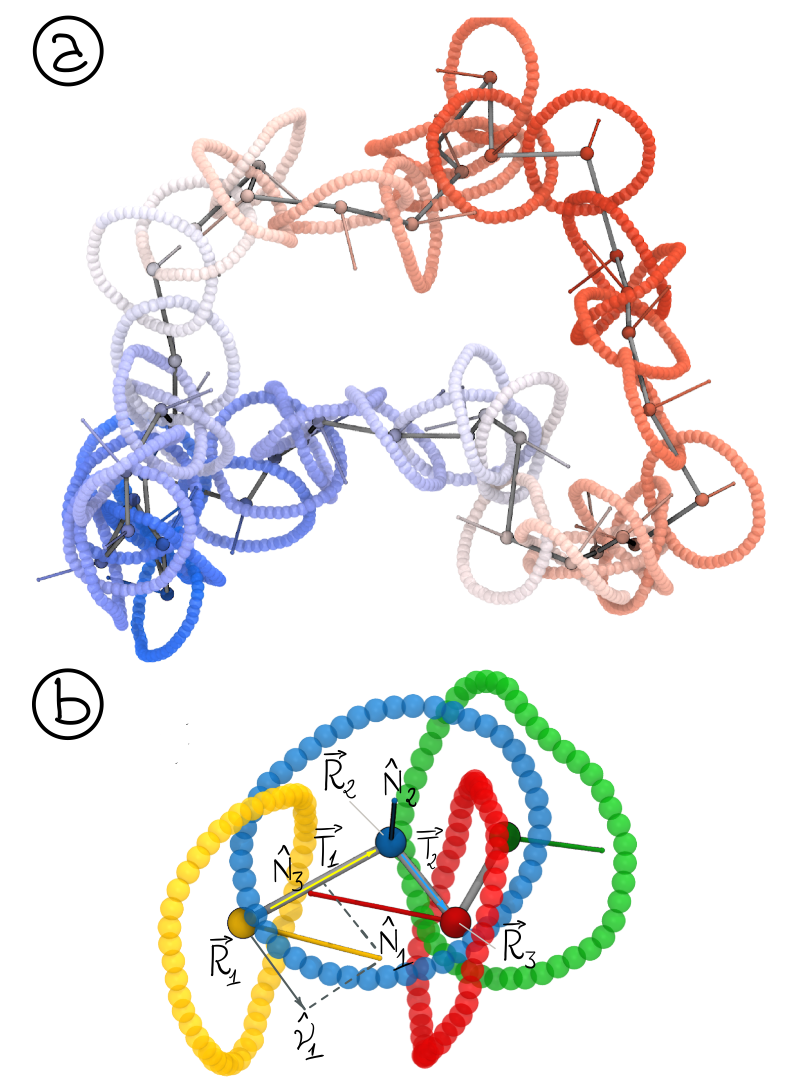}
    \caption{a) A conformation for a $n=40$, $n_{tr}=10$, $m=48$ catanenane superimposed to its coarse-grained backbone. The color map highlights the sequence of the elementary rings along the backbone. b)  Coarse-grained quantities for a small set of 4 subsequent rings. For ring 1, we also show the normal to the ribbon $\mathbf{\nu_1}$, obtained by subtracting from $\mathbf{\hat{N}_1}$ its projection along the bond $\mathbf{T_1}$.  }
    \label{fig:coarse-grain}
\end{figure}

In order to define the twist and writhe of the polycatenane, we need to map its configuration to that of a ribbon, using the coarse-graining strategy defined below.

\subsubsection{Coarse-graining the rings}
The large persistence length of our rings guarantees that their bending fluctuations are small, keeping them almost planar. This property allows us to coarse-grain the rings by identifying them with their center of mass and their normal, as shown in Fig.~\ref{fig:coarse-grain}. Specifically, for any index $k \in \{1,\ldots,n\}$ we identify the $k$-esim ring with the position of its center of mass, $\mathbf{R}_k$, and its normal versor $\mathbf{\hat{N}}_k$. The normal vector is defined ,using the indexing of the rings and following the the right-hand rule, as
\begin{equation*}
    \mathbf{N}_k = \frac{4}{m}\sum_{i=1}^{m/4} (\mathbf{r}_{i+m/4} - \mathbf{R}_k)\times(\mathbf{r}_{i} - \mathbf{R}_k),
\end{equation*}
and $\mathbf{\hat{N}}_k = \frac{\mathbf{N}_k}{|\mathbf{N}_k|}$.
Note that in general  an unoriented rigid ring corresponds to a  dyad and not a to vector, as the disc whose ring is the boundary  does not have an head or tail  face. The above definition of the normal versor is however possible because each elementary ring can be oriented in the reference conformation (the planar circle) as described in the main text, and this provides an ordering of the beads.    
Applying this coarse-graining procedure, we can thus map each  the ``microscopic" conformation of the polycatenane $\Gamma$ into the coarse grained representation $\Gamma_{cg} = \{(\mathbf{R}_1, \mathbf{\hat{N}}_1),\ldots,(\mathbf{R}_n, \mathbf{\hat{N}}_n) \}$.

\subsubsection{Twist and writhe}
By using  the normal versors and the positions, $\mathbf{R}_i$ we can define  the Twist and Writhe of the coarse grained representation, following the procedure described in Klenin and Langowski in~\cite{klenin2000computation}. To do so, we define the tangent to the polycatenane as the bond vector joining the centers of mass of two subsequent rings: $\mathbf{T}_k = \mathbf{R}_{k+1} -\mathbf{R}_k$ for $k<n$, and $\mathbf{T}_n = \mathbf{R}_1 - \mathbf{R}_n$, where $\mathbf{R}_k$ is the CoM of ring $k$. We recall that in the C\v{a}lug\v{a}reanu-White-Fuller theorem the normal and tangent to the ribbon are orthogonal to each other at any point. Thus, we define the normal versor to the ribbon, $\mathbf{\hat{\nu}_k}$ for the ring $k$ as:
\begin{equation}
    \mathbf{\hat{\nu}}_k = \frac{\mathbf{\hat{N}}_k - (\mathbf{\hat{N}}_k\cdot \mathbf{T}_k)\mathbf{T}_k}{||\mathbf{\hat{N}}_k - (\mathbf{\hat{N}}_k\cdot \mathbf{T}_k)\mathbf{T}_k||}\,.
\end{equation}

We  define the twist of a polycatenane as:
\begin{equation}\label{eq:twist}
    Tw(\Gamma) \equiv Tw(\Gamma_{cg}) =  \frac{1}{2\pi}\sum_{k=1}^n \phi_i,
\end{equation}
with $\phi_i = \alpha_i +\gamma_i$ where $\alpha_i$ is the angle between $\mathbf{\hat{\nu}_i}$ and the Frenet normal $\mathbf{\hat{B}_i}=\frac{\mathbf{T}_i\times\mathbf{T}_{i+1}}{||\mathbf{T}_i\times\mathbf{T}_{i+1}||}$, and $\gamma_i$ is the angle between $\mathbf{\hat{B}}_i$ and $\mathbf{\hat{\nu}_{i+1}}$. Note that $\phi_i$, $\alpha_i$, and $\gamma_i$ are  defined in the range $[-\pi, \pi]$ where the sign is determined by following the standard right-hand convention.

The writhe is defined as in ~\cite{klenin2000computation} by summing over  the solid angles $\Omega_{i,j}$ identified by bond vectors $\mathbf{T}_i$ and $\mathbf{T}_j$.
\begin{equation}\label{eq:writhe}
    Wr(\Gamma) \equiv Wr(\Gamma_{cg}) = \frac{1}{2\pi}\sum_{i=2}^n \sum_{j<i}^n \Omega_{ij}.
\end{equation}
If we call 1 and 2 the starting and ending point of the tangent vector $\mathbf{T}_i$ and 3,4 the ends of vector $\mathbf{T}_j$, we can define the vectors $\mathbf{r}_{12}$ joining points 1 and 2, $\mathbf{r}_{34}$ joining points 3 and 4, etc. Introducing the vectors
\begin{align*}
\mathbf{n}_1 = \frac{\mathbf{r}_{13}\times\mathbf{r}_{14}}{\mathbf{r}_{13}\times\mathbf{r}_{14}}, \qquad \mathbf{n}_2 = \frac{\mathbf{r}_{14}\times\mathbf{r}_{24}}{\mathbf{r}_{14}\times\mathbf{r}_{24}}, \\
\mathbf{n}_3 = \frac{\mathbf{r}_{24}\times\mathbf{r}_{23}}{\mathbf{r}_{24}\times\mathbf{r}_{23}}, \qquad
\mathbf{n}_4 = \frac{\mathbf{r}_{23}\times\mathbf{r}_{13}}{\mathbf{r}_{23}\times\mathbf{r}_{13}},
\end{align*}
we can finally calculate:
\begin{multline}
\Omega_{ij} = \arcsin{(\mathbf{n}_1\cdot\mathbf{n}_2)}+\arcsin{(\mathbf{n}_2\cdot\mathbf{n}_3)}+\\ +\arcsin{(\mathbf{n}_3\cdot\mathbf{n}_4)}+\arcsin{(\mathbf{n}_4\cdot\mathbf{n}_1)}.
\end{multline}
\subsubsection{Normal-normal correlation functions}
We define the normal-normal correlation function $C(d)$ as the averaged scalar product between two ring normals separated by $d$ "bonds":
\begin{equation}
    C_n(d) = \biggl<\frac{1}{n}\sum_{i=1}^n\mathbf{\hat{N}}_i\cdot\mathbf{\hat{N}}_{i+d}\biggr>_\Gamma,
\end{equation}
where the average runs over the independent conformations $\Gamma$ sampled by the Langevin dynamics. 
We note  that $C_n(d)$ measures the correlation between the orientation of the normals of the elementary rings, not of the ribbon. As the rings can orient themselves with their normals almost parallel to the backbone of the polycatenane, $C_n(d)$ capture more information than the correlation between the versors $\mathbf{\hat{\nu}}$ does


\end{document}